\documentclass[aps,preprint]{revtex4} 
\usepackage[dvips]{graphicx} 

\newcommand{\on}{\mbox{${\cal O}$(N)}}
\newcommand{\map}{\mbox{$^2\Delta\left[^1D\right]$}}
\newcommand{\hten}{\mbox{(H-H)$_5$}}
\newcommand{\hfour}{\mbox{(H-H)$_2$}}
\newcommand{\htwo}{\mbox{(H-H)}}
\newcommand{\ald}[1]{\mbox{HOC-{#1}}}
\newcommand{\Esub}{$E_{corr}^{HOC}$}
\newcommand{\dEabs}{$\left|\delta E_{corr}\right|$}

\newcommand{\figlabel}[1]{
\vspace{\stretch{5}}
 \\Figure #1\\Janesko et. al.\\Journal of Chemical Physics
\clearpage
\newpage
}

\begin{document} 

\title{Using molecular similarity to
  construct accurate semiempirical electronic structure theories}
  \author{Benjamin G. Janesko} \author{David Yaron}
  \email{yaron@cmu.edu} \affiliation{Department of Chemistry, Carnegie
  Mellon University,Pittsburgh, PA 15213}

\begin{abstract} 
{\em Ab initio} electronic structure methods give accurate results for
small systems, but do not scale well to large systems. Chemical
insight tells us that molecular functional groups will behave
approximately the same way in all molecules, large or small. This
molecular similarity is exploited in semiempirical methods, which
couple simple electronic structure theories with parameters for the
transferable characteristics of functional groups.
We propose that high-level calculations on small molecules provide a
rich source of parametrization data. In principle, we can select a
functional group, generate a large amount of {\em ab initio} data on
the group in various small-molecule environments, and "mine" this data
to build a sophisticated model for the group's behavior in large
environments. This work details such a model for electron correlation:
a semiempirical, subsystem-based correlation functional that predicts
a subsystem's two-electron density matrix as a functional of its one-electron
density matrix. This model is demonstrated on two small systems: chains of
linear, minimal-basis \hten , treated as a sum of four overlapping
\hfour\ subsystems; and the aldehyde group of a set of \ald{R}
molecules. The results provide an initial demonstration of the
feasibility of the approach.
\end{abstract} 

\maketitle 

\clearpage
\newpage

\section{Introduction}
\label{sec:intro}
Canonical {\em ab initio} electronic structure methods provide highly
accurate electronic structures for small systems of {\mbox{${\cal O}(10)$}}
atoms. However, these methods are too computationally intensive to
apply to large systems.  The formal scaling of computational effort
for {\em ab initio} calculations on an $N$-electron system ranges from
{\mbox{${\cal O}(N^3)$}} for Hartree theory
, to ${\cal O}(N^5)$ for MP2, to ${\cal
O}(e^N)$ for the exact, full-configuration-interaction (full-CI) solution~\cite{note1}.  {\em
Ab initio} (``first principles'') calculations always begin with a
minimal amount of information about the system (e.g. an initial
geometry and a basis set), determining practically all of the system's
features at runtime.

The computational effort of {\em ab initio} calculations can be
mitigated using two physically-motivated approximations: \on\ and
semiempirical approximations.

\on\ approximations are based on the principle of
nearsightedness~\cite{kohn_1rdm}, which states that the interactions
between parts of a molecule are largely {\em local} in character. (A
discussion of nearsightedness can be found in Ref.~\cite{goedecker}.)
\on\ approximations have been developed for every part of an {\em ab
initio} calculation, from fast multipole methods for Coulomb
effects~\cite{fmm,fmm2,fmm3} to divide-and-conquer~\cite{dac_dft} and
other~\cite{goedecker, conquest,mfcc} methods for self-consistent field
(SCF) calculations, to treatments of electron
correlation~\cite{ps1,pulay1,werner1,mhg1,as_ccd,werner3,mhg3,lmj,bart}. A
schematic of a nearsightedness-based approximation as outlined in
Ref.~\cite{kohn_1rdm} is shown in Fig.~\ref{fig:lrdm}.

Semiempirical approximations are based on the principle of molecular
similarity: that the properties of atoms and functional groups are
largely conserved in different molecules.  This principle formalizes
the chemical insights that methyl groups are relatively small
and nonpolar, halides are electron-withdrawing, and so
forth. {\em Ab initio} calculations spend much of their time in
re-calculating the transferable characteristics of functional groups.
Semiempirical approximations replace the {\em ab initio} Hamiltonian
with a simpler model Hamiltonian, which contains parameters that
capture the transferable characteristics of functional
groups. Examples of these parameters include force constants in
molecular mechanics~\cite{mm} or Hamiltonian matrix elements in
semiempirical quantum-mechanical approximations~\cite{indo,am1}.

One of the benefits of \on\ methods is their controllability. \on\
approximations yield well-defined changes in accuracy and
computational effort. The decision to use an \on\ approximation can be
made {\em a priori} based on the size of the system of
interest~\cite{goedecker}.

Unfortunately, semiempirical methods usually involve a significant trade-off between
computational effort and accuracy. Semiempirical methods are much less
accurate than {\em ab initio} methods for many systems.  This has led
to the widespread use of ``hybrid'' QM/MM methods, a
nearsightedness-based tradeoff between ab initio accuracy and
semiempirical speed~\cite{qmmm1a,qmmm2}. Our goal is to systematically
improve semiempirical theory.

Most existing semiempirical methods are based on models that were
designed to be parametrized to experimental data. Though many
semiempirical methods are now parametrized using {\em ab initio}
results (e.g Refs.~\cite{fromab_4, fromab_1, fromab_2, fromab_3}), we
believe that the existing methods may not take full advantage of the
possibilities inherent in {\em ab initio} parametrization. {\em Ab
initio} calculations on small molecules can give orders of magnitude
more parametrization data than can be readily obtained from
experiment. They also yield information that is more directly relevant
to a semiempirical model's parameters.

Nearsightedness and molecular similarity suggest that we can model
large systems as the sum of contributions from different functional
groups. This implies that a sufficiently rich data set of a functional
group in small molecules will contain {\em all} information needed to
describe the functional group in molecules of arbitrary size. Our
overall approach is to generate rich data sets on the behavior of
functional groups by doing a large number of highly accurate {\em ab
initio} calculations on the group in a set of small-molecule
environments. This paper investigates whether a semiempirical model
parametrized to this sort of small-molecule data can give {\em ab
initio} accuracy for larger molecules.

This approach is fairly general. It requires only that the
semiempirical model can describe a system as a sum of subsystem
contributions. For example, a semiempirical model that predicts the
amplitudes of delocalized wavefunctions would not be compatible with this approach.

The current work details our first implementation of this approach: a
semiempirical subsystem-based treatment of electron correlation. We
model the system in terms of its one- and two-electron density
matrices in an atomic orbital basis set
(Sec.~\ref{sec:dens}). Subsystem two-electron density matrices are combined
to model the two-electron density matrix of the entire system. This model was
chosen because it treats an important problem in contemporary electron
structure theory (electron correlation), and because the predicted
outputs (two-electron density matrices) are much easier to
obtain from {\em ab initio} calculation than from experiment.

\section{Methods}

\subsection{Semiempirical model for electron correlation}
\label{sec:dens}

Our semiempirical model treats electron correlation by predicting
subsystem two-electron density matrices as a functional of subsystem
one-electron density matrices. A system's one- and two-electron
density matrices $^1D,\ ^2D$ are obtained from its normalized
$N$-electron wavefunction $\left|\Phi\right\rangle$ as
\begin{equation} 
  \label{eq:1rdm} 
   ^1D(a,b) = \left\langle\Phi\right|a^{\dagger}_a a_b\left|\Phi\right\rangle 
\end{equation} 
\begin{equation} 
  \label{eq:2rdm} 
   ^2D(ac,bd)= 1/2 \left\langle\Phi\right|a_{a}^{\dagger}
  a_{c}^{\dagger} a_{b} a_d\left|\Phi\right\rangle 
\end{equation} 
in second quantization with one-electron basis functions
$\{\left|\phi_a\right>\}$. For an N-electron system, the trace of
$^1D$ equals N and the trace of $^2D$ equals the number of unique
electron pairs, $1/2\ N(N-1)$. The electron-electron interaction
energy of a system ($E_2$) is obtained as the trace over the product
of the two-electron integrals and the two-electron density matrix 
\begin{eqnarray}
  \label{eq:E2}
  E_2 &=& \sum_{abcd} \left<ac|bd\right>\ ^2D(ac,bd) \\ 
\left<ac|bd\right> &\equiv& \int d{\mathbf r_1}  d{\mathbf r_2} \phi_a^*({\mathbf r_1}) \phi_c^*({\mathbf r_2}) \frac{1}{\left|{\mathbf r_2} - {\mathbf r_1}\right|} \phi_b({\mathbf r_1}) \phi_d({\mathbf r_2}) \nonumber
\end{eqnarray}
The electron density in real space is the
diagonal of the one-electron density matrix: $^1D({\mathbf r}) \equiv
\left<\Phi\right|a^{\dagger}_{{\mathbf r}}a_{\mathbf
r}\left|\Phi\right>$~\cite{note2}. $^1D$ and $^2D$ provide
a complete description of a system whose Hamiltonian contains only
one- and two-body interactions~\cite{parr}.

The two-electron density matrix $^2D$ obtained from $\left|\Phi\right\rangle$
can be expressed as a cumulant expansion~\cite{maz2, maz_c2}
\begin{eqnarray}
\label{eq:decompose} 
^2D(ac,bd)& = &1/2\ ^1D(a,b) \ ^1D(c,d) \\ 
&-&1/2\  ^1D(a,d) \ ^1D(b,c)\nonumber\\
&+& ^2\Delta(ac,bd)\nonumber
\end{eqnarray}
where the three terms on the right-hand side of Eq.~\ref{eq:decompose}
are denoted Coulomb, exchange, and correlation contributions to
$^2D$. The connected pair density matrix $^2\Delta$ is that part of $^2D$
that cannot be written as a simple function of $^1D$. The Coulomb and
exchange contributions to $^2D$ in Eq.~\ref{eq:decompose} are
well-approximated at the Hartree and Hartree-Fock levels of theory,
respectively. However, accurate {\em ab initio} treatment of the
connected pair density matrix $^2\Delta$ requires expensive high-level
methods.

Density functional theory (DFT) is a formally exact method for
treating a system of interacting electrons exclusively in terms of its
one-electron density~\cite{hk,ks,parr}. The electron-electron
interaction energy of Eq.~\ref{eq:E2} is treated as the sum of a
Coulomb term and an exchange-correlation correction $E_{XC}$, such
that the electrons move in a potential that is corrected by the
exchange-correlation potential
$ v_{XC}({\mathbf r}) = \delta(E_{XC}) /
\delta(^1D({\mathbf r}))$
. DFT is implemented by approximating $v_{XC}$ as a functional of
electron density: $v_{XC} = v_{XC}[^1D({\mathbf r})]$. (In Kohn-Sham DFT, the
kinetic energy is decomposed into the kinetic energy of the Kohn-Sham 
orbitals plus a density-dependent correction, which is incorporated
into $v_{XC}$ via e.g. adiabatic connection~\cite{parr}. Our
$v_{corr}$ functionals (Eq.~\ref{eq:vcorrpca}) do not include a
kinetic energy correction.)  Following Eq.~\ref{eq:E2} and
Eq.~\ref{eq:decompose}, $E_{XC}$ may be obtained as the trace over the
exchange and correlation contributions to the two-electron density matrix:
$E_{XC} = \sum \left<ac|bd\right> \left(-1/2\ ^1D(a,d)\ ^1D(b,c) +\
^2\Delta(ac,bd) \right)$. Thus, the correlation component
$v_{corr}[^1D]$ of a system's exchange-correlation functional can be
obtained from the first derivative of a functional that predicts a
system's connected pair density matrix $^2\Delta$ as a function of its
electron density matrix $^1D$:
\begin{eqnarray}
\label{eq:vcorrderiv}
 v_{corr}[^1D](a^{'},b^{'}) =
\sum_{abcd} \left<ac|bd\right> \\
\times  \delta\left(^2\Delta[^1D]\left(ac,bd\right)\right)/\delta(^1D(a^{'},b^{'})) \nonumber
\end{eqnarray}
Explicit treatments of $v_{XC}[^1D]$ in terms of the two-electron
density  include various analyses of the real-space
exchange-correlation hole~\cite{vxc_constraints,xchole_nucleii,xhole}.

In this work, we define the correlation energy $E_{corr}$ as the
expectation value of the connected pair density matrix: $E_{corr} = \sum
\left<ac|bd\right>\ ^2\Delta(ac, bd)$. Correlation energy can also be
defined as the difference in energies predicted by
configuration-interaction and Hartree-Fock calculations: $E_{corr} =
E_{CI} - E_{HF}$. The latter definition includes the effects of $^1D$
relaxation, e.g. the expectation value of $^1D_{CI} -\ ^1D_{HF}$. In
contrast, the former definition yields the correlation energy
corresponding to a single choice of $^1D$, and is therefore consistent
with the definition of $E_{corr}$ used in DFT and MP2 calculations.

Both $^1D$ and $^2\Delta$ can be treated using the nearsightedness
approximation.  Several nearsightedness-based treatments of $^1D$
exist, including divide-and-conquer methods that partition $^1D$ into
subsystem contributions as in Fig.~\ref{fig:lrdm}~\cite{goedecker,
dac_dft,dac_semM, dac_semY, gauss_dac_dft,
dac_dft_bias}. Nearsightedness-based treatments of $^2\Delta$ include
the \on\ treatments of electron correlation cited
previously~\cite{ps1,pulay1,werner1,mhg1,as_ccd,werner3,mhg3,lmj,bart}. We
recently developed the ``localized reduced density matrix'' (LRDM)
method~\cite{lrdm}, a divide-and-conquer style treatment of
$^2\Delta$. LRDM assembles a large system's atomic-orbital-basis
$^2\Delta$ from the results of {\em ab initio} calculations on
overlapping subsystems. Like other
divide-and-conquer methods, LRDM is non-variational.

In the current work, we use LRDM as a framework for semiempirical
subsystem-based approximations for DFT correlation functionals
$v_{corr}[^1D]$. We generate semiempirical functionals that predict
the matrix elements of a subsystem's $^2\Delta$ as a function of the
subsystem electron density matrix $^1D$: \map\ (see Eq.~\ref{eq:twodeltapca}
and Eq.~\ref{eq:vcorrpca} below). An approximate $^2\Delta$ for a
large system is obtained by combining subsystem \map\ predictions
using LRDM. Our results indicate that basis-set \map\
functionals can provide good results for multiple subsystem geometries
(Sec.~\ref{sec:results}).

LRDM can treat long-range correlations (dispersion interactions) by
doing {\em ab initio} calculations that include correlation in two
disjoint regions of a molecule~\cite{lrdm}. In the current work, we do
not model these long-range interactions. Therefore, our
subsystem-based $v_{corr}[^1D]$ functionals, like standard DFT $v_{corr}$
functionals~\cite{dft_vdw1,dft_vdw2}, cannot treat dispersion
interactions.

Our subsystem-based $v_{corr}[^1D]$ functionals are very different than the
standard DFT functionals derived from the homogeneous electron
gas~\cite{ks,parr,ldarev,gga1}. Other groups have developed
$v_{XC}[^1D]$ functionals that are
semiempirical~\cite{b3lyp,overbind}, subsystem-based~\cite{edge,vxsub}
or fitted to high-accuracy {\em ab initio} data~\cite{nnVxc,
abVxc_1,abVxc_2}, but to our knowledge the current method is unique in
combining semiempirical methods with a nearsighted,
molecular-similarity-based treatment of $^2\Delta$.

\subsection{Parametrization method}
\label{sec:param}

Our approach is to develop semiempirical models that are parametrized
using rich data sets of small-molecule {\em ab initio}
calculations. These rich data sets allow us to use data mining methods
in the parametrization stage. ``Data mining'' refers to computational
methods for analyzing large data sets and automatically extracting
previously unknown dependencies between the
data~\cite{databook}. Other data-mining treatments of electron
correlation include a neural-network exchange-correlation potential
fitted to data from many molecules~\cite{nnVxc}, and a model for the
correlation energy between pairs of widely separated, localized
electrons~\cite{nnEcorr}.

Data mining methods can determine two types of relationships between
data. The first is the system's dimensionality: which input and output
descriptors are most important for describing the data set. (Here,
``descriptor'' is a generic term for a type of data used by a
model. For example, the input and output descriptors of our \map\
functionals are $^1D(a,b)$ and $^2\Delta(ac,bd)$ matrix elements.)
The second type of relationship that can be determined by data mining
is the functional form of the [input descriptor]$\to$[output
descriptor] relation. In the current work, we assume a quadratic
input-output relation and focus on dimensional reduction.

A flowchart of the data-mining process for a functional group is as
follows.

\begin{enumerate}
\item Choose an initial set of input and output descriptors, and a fit
function to relate them. As discussed in Sec.~\ref{sec:intro},
subsystem-based models require input and output descriptors that
describe electronic structure in terms of local information
(e.g. electron densities). Since the models are meant for use
within semiempirical models, the input descriptors should be
obtainable from a simple approximate Hamiltonian (e.g. the DFT
Hamiltonian). The fit function can be anything from a polynomial fit
to a neural network.
\item Generate an initial data set of {\em ab initio} calculations on
the functional group in various small molecules and
environments. Extract the values of all input and output descriptors
for each point in the data set.
\item \label{item3} Split the data set into training and testing subsets.
\item \label{item4} Reduce the dimensionality of the data set, by
using (for example) principal component analysis to determine a few
combinations of descriptors that capture most of the variation in the
data set. The model will be parametrized on this
dimensionally-reduced input and output data.
\item Parametrize the model using the training subset of the
small-molecule data.
\item Test the model on the testing subset of the small-molecule data,
and on larger molecules.
\end{enumerate}

As stated above, our initial focus is on the dimensional reduction of
$^1D$ and $^2\Delta$ (step~\ref{item4}). For a system with $M$ basis
functions, $^1D$ and $^2\Delta$ have $1/2\ (M^2+M)-1$ and
$1/8\ (M^4+2M^3+5M^2+4M)$-1 degrees of freedom, respectively. Without
dimensional reduction, even a small subsystem (e.g. M = ${\cal
O}(10)$) has far too many output degrees of freedom for a \map\
functional to be useful. We use principal component analysis (PCA) to
decompose $^1D$ and $^2\Delta$ into a set of importance-weighted basis
functions. To illustrate, PCA on a set of subsystem electron density matrices
$\{^1D_{\sigma}\}$ describes each density matrix as
\begin{equation}
\label{eq:pca}
^1D_{\sigma}(a,b) =\ ^1D_{avg}(a,b) + \sum_i c_{\sigma i}\ ^1D_i(a,b)
\end{equation}
where $^1D_{avg}$ is the average electron density matrix, $^1D_i$ are the
principal components, and the standard deviation of the expansion
coefficients $c_{\sigma i}$, evaluated across the data points $\sigma$, decreases with
increasing $i$~\cite{databook}.

In this work, we use a quadratic function to predict the first few
(most important) $^2\Delta$ components from the first few $^1D$
components. A quadratic function is the lowest-order polynomial of
$^2\Delta[^1D]$ for which the associated correlation-energy functional
$v_{corr}[^1D]$ (Eq.~\ref{eq:vcorrderiv}) is not a constant. The \map\
functionals fit the first $C_2$ principal components of $^2\Delta$ as
a function of the first $C_1$ components of $^1D$ such that
\begin{eqnarray}
\label{eq:twodeltapca}
^2\Delta[^1D](ac,bd) = \ ^2\Delta_{avg}(ac,bd)  + \sum_{j}^{C_2}\{  ^2\Delta_j(ac,bd)\\ 
  \times (\alpha_j + \sum_{i}^{C_1} \left( \gamma_{ij} \left(^1D|^1D_i\right) + \sigma_{ij}
\left(^1D|^1D_i\right)^2\right))\}  \nonumber
\end{eqnarray}
where $^2\Delta_j$ are the principal components of
$^2\Delta$, $\left(^1D|^1D_i\right)$ is the projection of the argument
one-electron density matrix $^1D$ onto the $i$th principal component
\begin{equation}
\left(^1D|^1D_i\right) \equiv \sum_{ab} \left(^1D(a,b) - \ ^1D_{avg}(a,b)\right)\ ^1D_i(a,b) 
\end{equation}
and \{$\alpha_j, \gamma_{ij}, \sigma_{ij}$\} are fitted
parameters. Each component of the two-electron density matrix is fit
independently of the others.  The subsystem DFT correlation energy
operator $v_{corr}[^1D]$ is obtained from \map, following
Eq.~\ref{eq:vcorrderiv}, as
\begin{eqnarray}
\label{eq:vcorrpca}
(v_{corr}[^1D])(a^{'},b^{'})  =  \sum_{abcd}\left<ac|bd\right>
\sum_{j}^{C_2} \Big\{ ^2\Delta_j(ac,bd)
\nonumber\\ \times  \sum_i^{C_1} \ ^1D_i(a^{'}, b^{'}) \left(\gamma_{ij} + 2 \sigma_{ij}
\left(^1D|^1D_i\right)\right) \Big\}
\end{eqnarray}
The $v_{corr}[^1D]$ functional of a large system is obtained by
overlaying subsystem contributions as in LRDM
(Fig.~\ref{fig:lrdm}). The degree of dimensional reduction can be
seen by comparing the number of fitted components $C_1$ and $C_2$ to
the total number of degrees of freedom in $^1D$ and $^2\Delta$.

The approach discussed here can be used to construct many different
kinds of semiempirical model based on the choice of input and output
descriptors. For example, we have begun work on a semiempirical model
of core polarization in effective core potentials~\cite{ecp}. Here,
the input descriptors are the valence electron density matrix and
one-electron Hamiltonian, and the output descriptors are the core
electron density matrices. This work will be discussed in a
future publication.

The principal computational challenge of this approach is the steep
scaling of the amount of training set data required.  The \map\
functional of an M-orbital subsystem will have \mbox{I$ = {\cal
O}$(M$^2$)} $^1D_i$ input components~\cite{note3}. In general, a
function with I input components must be parametrized using ${\cal
O}(e^{\textrm{I}})$ data points~\cite{databook}. Because of this, we
have focused our initial work on proof-of-concept treatments for small
model systems in minimal basis sets. Successful application of our method to these
systems will justify expanding our focus to larger, chemically-interesting
systems and larger basis sets.

\subsection{Error Decomposition}
\label{sec:restype}

Our \map\ functionals contain three distinct approximations. The first
approximation is nearsightedness: the connected pair density matrix
$^2\Delta$ is assumed to be well-described by a decomposition into
overlapping subsystems. The second approximation is that each
subsystem $^2\Delta$ is assumed to be well-described by a relatively
small number of principal components ($C_2$ in
Eq.~\ref{eq:twodeltapca}). The third approximation is that each
subsystem $^2\Delta$ component is assumed to be well-described by the
$^1D$ functional in Eq.~\ref{eq:twodeltapca}.

We can isolate the effects of each of these assumptions using three
kinds of approximate connected pair density matrix (see
Table~\ref{tab:rho2types}). The first approximate connected pair
density matrix is the ``exact subsystem'' connected pair density
matrix: $^2\Delta_{xsub}$. This is obtained by projecting the correct
$^2\Delta$ onto the overlapping subsystems, and setting to zero all
matrix elements that are not contained within a subsystem. The second
is the principal component reduction connected pair density matrix:
$^2\Delta_{PCA}$. This is obtained by projecting the subsystem blocks
of $^2\Delta_{xsub}$ onto the $^2\Delta$ components that are fitted by
the subsystem \map\ functionals~\cite{note4}. The third is the
connected pair density matrix obtained using the subsystem \map\
functional and the correct one-electron density matrix:
$^2\Delta[^1D_{exact}]$. Table~\ref{tab:rho2types} summarizes the
approximations used in generating these density matrices.

\section{Results}
\label{sec:results}

The remainder of this paper details demonstrations of our
semiempirical subsystem-based \map\ functionals.  We begin by
demonstrating \map\ functionals for a linear dimerized chain of
minimal-basis hydrogen atoms \hten, since the functional predictions
can be readily compared to full-CI. Then, we demonstrate that a \map\
functional for the aldehyde group, parametrized to data from a set of
small \ald{R} molecules, can extrapolate to R groups outside of the
training set. All {\em ab initio} calculations were performed using a
modified version of the GAMESS electronic structure
program~\cite{GAMESS}.

\subsection{\hfour\ and \hten\ systems}
\label{sec:generate}
The first system is linear minimal-basis \hten. This system is treated
as a sum of four overlapping \hfour\ subsystems. We model its
correlation energy by parametrizing a \hfour\ \map\ functional to
data on isolated \hfour\ molecules, and combining the \hfour\ \map\ predictions
using LRDM~\cite{note5}. The functionals are parametrized to, and tested on,
full-CI {\em ab initio} calculations.

We generated data for both variable- and fixed-geometry molecules,
yielding the four data sets in Table~\ref{tab:hchaindata}.  Each
molecule was electrostatically perturbed by randomly placing
fractional charges ($|$charge$|\leq 1$) into a $6 \AA \times 6 \AA
\times (\textrm{molecule length} + 4 \AA)$ box around the molecule,
with a minimum point charge - atom separation of 1.2 \AA. Variable
geometry systems had each bond length set randomly within the ranges
in Table~\ref{tab:hchaindata}.

The \hfour\ \map\ functionals were parametrized using half of the
\hfour\ data as a training set (see item~\ref{item3} in the flowchart
of Sec.~\ref{sec:param}). Separate functionals were parametrized for
the variable- and fixed-geometry systems. The numbers of principal
components included in the \map\ functionals ($C_1$ and $C_2$ in
Table~\ref{tab:hchaindata}) were selected to give good results for
both $^2\Delta[^1D_{exact}]$ and $^2\Delta[^1D_{DFT}]$ (see below). The principal
component analyses were a significant dimensional reduction, as the
$^1D$ and $^2\Delta$ of \hfour\ contain 9 and $59$
degrees of freedom, respectively.

\subsection{Modeling \hfour\ using \hfour\ \map\ functionals}
\label{sec:hfourres}

The first test of the \hfour\ \map\ functionals is how well they can
predict the \hfour\ $^2\Delta$ given the correct full-CI electron
density matrix $^1D_{exact}$.  Fig.~\ref{fig:res1} plots predicted
vs. real $E_{corr}$ for the \hfour\ systems. Table~\ref{tab:res1}
presents \dEabs\ errors averaged over the training- and testing-set
data. Here, \dEabs\ is the absolute error in the predicted correlation
energy $E_{corr} = \sum_{abcd} \left<ac|bd\right>\ ^2\Delta(ac,bd)$
(see Eq.~\ref{eq:E2}). The \map\ $E_{corr}$ predictions are compared
to MP2.

The results are quite encouraging. The \map\ functionals are better
than MP2 at predicting the average value of the correlation energy:
the mean absolute errors in $E_{corr}$ from MP2 are 40 and 150 times
as large as the error for the \map\ functionals (variable and fixed
geometry, respectively). \map\ functionals are also better than MP2 at
predicting the variation of the correlation energy across the data
set. This can be seen in Fig.~\ref{fig:res1}: the slope of the
predicted vs. real $E_{corr}$ values is very small for the MP2
predictions but close to 1 for the \map\ functionals. Despite its low
scatter, MP2 does not capture either the value or the variation in the
correlation energy.

The scatter in the \map\ predictions for the fixed-geometry system can
be reduced by parametrizing a \map\ functional that uses more
principal components. We parametrized a \map\ functional for
fixed-geometry \hfour\ that includes seven $^1D$ and eight $^2\Delta$
principal components ($C_1=7,\ C_2=8$ in
Eq.~\ref{eq:twodeltapca}). This functional gives an R$^2$ between real
and predicted $E_{corr}$ of 0.990, comparable to the 0.991 value for
MP2 and better than the 0.890 value in Fig.~\ref{fig:res1}. For this
functional, the average (standard deviation) $^2\Delta[^1D_{exact}]$
\dEabs\ values are 0.11 (0.29) mH
for the testing-set data.

A comparison of the $^2\Delta_{PCA}$ and $^2\Delta[^1D_{exact}]$
errors in Table~\ref{tab:res1} shows that most of the error in the
variable-geometry system is due to dimensional reduction of
$^2\Delta$, as the $^2\Delta_{PCA}$ errors are almost as large as the
corresponding $^2\Delta[^1D_{exact}]$ errors. In contrast, the error
in the fixed-geometry system is more evenly partitioned between
dimensional reduction of $^2\Delta$ and prediction of $^2\Delta$ from $^1D$. 

The training- and testing-set errors are reasonably close to each
other, indicating that the functionals are not over-fitted. We tested
a second measure of the predicted $^2\Delta$, the sum of absolute
errors in the predicted $^2\Delta$ matrix elements. These errors were
fairly well-correlated with the \dEabs\ errors presented above (data
not shown).

The results in Table~\ref{tab:res1} and Fig.~\ref{fig:res1} show that
the constant $^2\Delta$ returned by $^2\Delta[^1D_{avg}]$ is a
surprisingly good approximation for the variable-geometry
systems. This is encouraging, as it suggests that even the most
primitive \map\ functional (e.g. a constant $^2\Delta$) can work
reasonably well for multiple subsystem geometries.  Our \map\
functionals all improve upon this primitive functional, as all
$^2\Delta[^1D_{exact}]$ errors are lower than the corresponding
$^2\Delta[^1D_{avg}]$ errors. As expected, $^2\Delta[^1D_{avg}]$
predicts a constant correlation energy for the fixed-geometry systems
(Fig.~\ref{fig:res1}).

\subsection{Modeling \hten\ using \hfour\ \map\ functionals}
\label{sec:htenres}

The results in Fig.~\ref{fig:res1} and Table~\ref{tab:res1}
demonstrate that the \hfour\ \map\ functionals give good $^2\Delta$
predictions for \hfour. Given this, we investigate whether four copies
of an \hfour\ \map\ functional, combined using LRDM, will suffice to
describe correlation effects in \hten. Using the \hfour\ \map\
functional on \hten\ tests whether the fundamental assumptions of
nearsightedness and molecular similarity, and our implementation of
these approximations, are correct for the \hten\ model system.
Fig. ~\ref{fig:res2} and Table~\ref{tab:res2} present data for \hten\
systems, using the notation of Fig.~\ref{fig:res1} and
Table~\ref{tab:res1}.

The \hten\ results are also encouraging. Four copies of an \hfour\
\map\ functional, combined via LRDM, are better than MP2 at describing
the mean and variation of $E_{corr}$ for the \hten\ system. The mean
absolute $E_{corr}$ errors for MP2 are 60 and 90 times the values for
\map\ functionals (variable and fixed geometry,
respectively). Fig. ~\ref{fig:res2} shows that our method does better
than MP2 at capturing the variation in $E_{corr}$ across the data set,
with a predicted vs. real $E_{corr}$ whose slope is very small for MP2
but near one for our method.

For the variable-geometry systems, the \map\ functionals describe the
\hten\ data to about the same level of accuracy (per atom) as the
\hfour\ data. The average \hten\ \dEabs\ are about $10/4=2.5$ times as
large as the corresponding \hfour\ values. For example, the average
$^2\Delta[^1D_{exact}]$ error is 1.91 mH for variable-geometry \hfour\
and 3.43 mH for variable-geometry \hten.

For the fixed-geometry systems, the \map\ functionals do {\em not}
describe the \hten\ data to the same level of accuracy as the \hfour\
data: the average \hten\ \dEabs\ are about five times as large as the
corresponding \hfour\ values. This error is not due to the subsystem
decomposition: the average \dEabs\ of $^2\Delta_{xsub}$ is only 0.02
mH (Table~\ref{tab:res2}). We suggest that the long-range order in the
fixed-geometry \hten\ leads to an intrinsic difference between the
environments experienced by an isolated \hfour\ vs. an \hfour\
embedded in \hten. Better \hfour\ \map\ functionals for the
fixed-geometry systems could perhaps be generated by using cyclic
boundary conditions in the \hfour\ data. Evidence for this conclusion
is discussed in the Supporting Information.

The predictions of a semiempirical model should not depend on the
choice of training set used to parametrize the model. We parametrized
\hfour\ \map\ functionals using multiple choices of training
set. Results are discussed in the Supporting Information. As expected,
the functionals have only a weak dependence on training set choice.

\subsection{DFT calculations with \map\ functionals}

The above results test the \map\ functional's predictions given the
correct electron density matrix $^1D_{exact}$. However, the functionals are
intended for use in density functional theory (Sec.~\ref{sec:dens})
where $^1D_{exact}$ is not known in advance. We have implemented two
methods for using the functionals. The first is DFT with exact
exchange and the \map\ correlation functional of
Eq.~\ref{eq:vcorrpca}, referred to as $^2\Delta[^1D_{DFT}]$. The
second method, like MP2, is a one-step post-Hartree-Fock prediction of
$E_{corr}$. Here, the correct electron density matrix $^1D_{exact}$ is
approximated as the Hartree-Fock electron density matrix $^1D_{HF}$, and the
correlation energy is obtained non-self-consistently from
$^2\Delta[^1D_{HF}]$.

Table~\ref{tab:dft1} presents \dEabs\ values for $^2\Delta[^1D_{DFT}]$
and $^2\Delta[^1D_{HF}]$ on the fixed- and variable-geometry \hfour\
and \hten\ systems, for a single choice of training set. Predicted
vs. real $E_{corr}$ for the \hten\ systems are plotted in
Fig.~\ref{fig:dft2}.

The $^2\Delta[^1D_{DFT}]$ calculations do a fairly good job of
predicting the average and variation in $^2\Delta$ and $E_{corr}$: the
average and standard deviations in \dEabs\ are much better than MP2,
and the standard deviations in \dEabs\ are generally smaller than the
primitive $^2\Delta[^1D_{avg}]$ functional (see Tables~\ref{tab:res1}
and~\ref{tab:res2}). These results are encouraging, given that our
$v_{corr}[^1D]$ functional is a simple linear function of $^1D$ (see
Eq.~\ref{eq:vcorrpca}). The results from the \hten\ systems are
especially encouraging: four identical, overlapping \hfour\
$v_{corr}[^1D]$ functionals give a reasonable prediction for the
$v_{corr}[^1D]$ of \hten. Fig.~\ref{fig:dft2} shows that the
relatively large average errors in the $^2\Delta[^1D_{DFT}]$ \dEabs\
are mostly systematic error. Better $^2\Delta[^1D_{DFT}]$ results
could perhaps be generated using a more sophisticated (nonlinear)
function (see flowchart, Sec.~\ref{sec:param}). The errors in the
non-self-consistent $^2\Delta[^1D_{HF}]$ calculations are somewhat
higher than the self-consistent $^2\Delta[^1D_{DFT}]$
calculations. This is reasonable, especially given that $^1D_{HF}$ is
not necessarily a good approximation for $^1D_{exact}$.

When \map\ functionals are combined with exact exchange, the
$^2\Delta[^1D_{HF}]$ and $^2\Delta[^1D_{DFT}]$ calculations give a
fairly large, systematic under-estimate of the total energy. This is
partly due to a difference between the $^1D$ obtained using full-CI
and HF theory on minimal-basis \hten. For a closed-shell N-electron
system, the combined trace of the exchange and correlation parts of
$^2D$ (Eq.~\ref{eq:decompose}) equals $-1/2\ N$~\cite{parr}. Full-CI
calculations on \hten\ give $^2\Delta$ with a trace $<0$ and an exact
exchange pair density matrix $^2D_{X}(ac,bd) = -1/2\ ^1D(a,d)\
^1D(b,c)$ with a trace less than $-1/2\ N$. Thus, all of the \hfour\
\map\ functionals return a $^2\Delta$ with a negative trace. However,
{\em ab initio} methods that return a single-determinant wavefunction
(e.g. HF or KS-DFT theory) give a closed-shell $^2D_{X}$ with a trace
equal to $-1/2\ N$. Thus, for example, the approximate two-electron
density matrix returned by our non-self-consistent method
$^2D(ac,bd) = 1/2\ ^1D_{HF}(a,c)\ ^1D_{HF}(b,d) - 1/2\
^1D_{HF}(a,d)\ ^1D_{HF}(b,c) +\ ^2\Delta[^1D_{HF}](ac,bd)$ will always
have a trace less than the correct value ($1/2\ N(N-1)$, see
Sec.~\ref{sec:dens}). This leads to a systematic under-estimate of the
number of electron pairs in the system and the electron-electron
interaction energy. One way to correct this is by renormalizing the
exact-exchange $^2D_{X}$ obtained from $^1D_{HF}$ or $^1D_{DFT}$, such
that the final predicted $^2D$ has the correct trace. This is
analogous to the use of a fraction of exact exchange in ``hybrid'' DFT
functionals such as B3LYP~\cite{b3lyp}. This significantly improves
the total energies: for example, the average (standard deviation)
total energy error for variable-geometry \hten\ is \mbox{114.71
(13.50) mH} for uncorrected Hartree-Fock calculations and \mbox{-71.52
(7.26) mH} and \mbox{-160.88 (19.67) mH} for $^2\Delta[^1D_{HF}]$ with
and without renormalization of $^2D_{X}$.

\subsection{Substituted aldehydes}
\label{sec:ald}

The assumption of molecular similarity implies that a \map\ functional
for the aldehyde group of \ald{R} molecules should be able to
extrapolate to R groups outside of its training set. We tested this
assumption by parametrizing aldehyde \map\ functionals using
minimal-basis (STO-3G) \ald{R} molecules with six different R groups:
H, F, OH, CH$_3$, Cl, and OCH$_3$. Six different \map\ functionals
were generated from this data. Each was trained on a data set that
excluded data from one of the six R groups, and included half of the
data from the other five groups. The functionals were tested for their
ability to accurately model the aldehyde for both the five kinds of
\ald{R} molecules in the training set and the R group excluded from the
training set.

Details of the calculation are as follows. The {\em ab initio} data
set contained 250 calculations for each of the six kinds of \ald{R}
molecules. Each calculation had random geometric~\cite{note6} and
electrostatic~\cite{note7} perturbations similar to those in the
variable-geometry \hten\ chains above. {\em Ab initio} calculations
were performed using MP2, as the different-sized \ald{R} groups
required a size-consistent method and full-CI was prohibitively
expensive. The aldehyde \map\ functionals were fitted to the MP2
$^2\Delta$ and the relaxed $^1D$~\cite{relax1} of the aldehyde
group. The aldehyde functionals' performance was characterized by
their ability to reproduce the ``aldehyde correlation energy'' defined
as \mbox{$E_{corr}^{HOC} = \sum \left<ac|bd\right>\ ^2\Delta(ac, bd)\
;\ \{abcd\}\in \textrm{HOC}$}. All functionals used 40 $^1D$ and 30
$^2\Delta$ principal components. This was a significant dimensional
reduction, as the aldehyde $^1D$ and $^2\Delta$ contain 65 and $\sim
2200$ degrees of freedom. Results from the six functionals are
presented in Table~\ref{tab:ald1}. A plot of the extrapolation results is
in Fig.~\ref{fig:aldplot2}.

In general, the results are quite good.  $^2\Delta[^1D_{exact}]$
errors for the R groups in the training sets(Table~\ref{tab:ald1},
upper panel, off-diagonals) are small compared to both the average
\Esub\ (-139.60 mH) and the standard deviation in \Esub\ (13.30
mH). Most of the extrapolations are also good, with
$^2\Delta[^1D_{exact}]$ errors that are uniformly smaller than the
corresponding $^2\Delta[^1D_{avg}]$ errors (diagonals of
Table~\ref{tab:ald1}, compare upper and lower panels). The
$^2\Delta[^1D_{avg}]$ energy errors are fairly good, as in the
variable-geometry \hten\ systems, providing further evidence that the
primitive, constant-$^2\Delta$ functional works rather well for
multiple geometries (see Sec.~\ref{sec:hfourres}).

\section{Discussion}

Nearsightedness and molecular similarity suggest that a rich data set
of {\em ab initio} calculations on a functional group in various small
molecules contains sufficient information to describe the functional
group's behavior in large molecules. Here we explore new methods for
generating semiempirical electronic structure models that are
parametrized to such data sets. In particular, we consider a
semiempirical, subsystem-based model of electron correlation. This
model predicts the connected pair density matrix $^2\Delta$ of molecular
subsystems as a functional of the subsystem electron density matrix
$^1D$. Subsystem $^2\Delta$ predictions are combined using a
previously-developed divide-and-conquer-style treatment of the
atomic-orbital-basis $^2\Delta$ (LRDM). The $^2\Delta$ functionals are
used to obtain correlation-energy functionals for density functional
theory (Eq.~\ref{eq:vcorrpca}).  The method is tested on chains of
minimal-basis \hten, which was treated as a system of four identical
and overlapping \hfour\ subsystems. The extrapolation abilities of
\map\ functionals are tested on \ald{R} molecules.

The \hten\ chain results demonstrate that the model works well for
these simple systems. The results in Fig.~\ref{fig:res1} and
Table~\ref{tab:res1} show that \map\ functionals fitted to {\em ab
initio} data on \hfour\ can reproduce the \hfour\ $^2\Delta$ given the
correct electron density matrix $^1D_{exact}$.  The $^2\Delta$ of \hten\
systems can be modeled quite well using four overlaid \hfour\ \map\
functionals, as shown by the data in Fig.~\ref{fig:res2} and
Table~\ref{tab:res2}. Fig.~\ref{fig:dft2} and Table~\ref{tab:dft1}
show that the \map\ functionals work reasonably well as DFT
correlation functionals. These results are especially encouraging
given the simple, linear form of the $v_{corr}[^1D]$ functionals
(Eq.~\ref{eq:vcorrpca}). The subsystem \map\ functionals can extrapolate
to molecules outside of the training set, as demonstrated by the
\ald{R} results in Sec.~\ref{sec:ald}.

An interesting finding is that dimensional reduction of subsystem
$^2\Delta$ seems to be a reasonable approximation. All of the \map\
functionals had significant reductions in the dimensionality of
$^2\Delta$. This suggests that real molecular environments only
explore a fraction of the total degrees of freedom in a functional
group's $^2\Delta$. This dimensional reduction may be useful for other
models of electron correlation.  Our results also suggest that, for
these systems, simple quadratic functions are a fairly good model for
the input:output relation of the dimensionally reduced data.

It is also interesting that subsystem \map\ functionals that are
defined in a basis set can be used for multiple subsystem
geometries. For both hydrogen chains and \ald{R} molecules, a single
\map\ functional provided good $^2\Delta$ predictions for a fairly
wide range of different geometries.  Real-space \map\ functionals may
be more general than those presented here. However, the success of the
basis-set functionals is encouraging.

The utility of our semiempirical, subsystem-based treatment of
electron correlation relies upon its ability to provide high-accuracy
treatments of systems that are too large for current {\em ab initio}
methods. In particular, we would like to perform high-accuracy DFT
calculations on polypeptides, using \map\ or $^2D_{XC}[^1D]$
functionals for the twenty most common amino acids. In order to
provide an accurate treatment of electron correlation, each of the
twenty functionals would be parametrized to high-level {\em ab initio}
calculations in a large basis set
(e.g. CCSD(T)/cc-pVTZ)~\cite{ccbasis}.  These calculations would place
each amino acid in a large number of small-molecule environments. The
environments could include the amino acid in various 2- or 3-residue
polypeptides, or with capping groups as in Ref.~\cite{mfcc}.

As stated above, the principle challenge to reaching this goal is the
computational expense of parametrizing subsystem \map\ or
$^2D_{XC}[^1D]$ functionals for moderately-sized subsystems in a large
basis set. Parametrizing a \map\ functional for a group with $M$
basis functions requires ${\cal O}(exp(M^2))$ data points. In the
current proof-of-concept work, we studied small subsystems in a
minimal basis set, so that new \map\ functionals could be parametrized
and tested in a relatively short time. Data sets for the systems
studied here, minimal-basis \hten\ and \ald{R}, could typically be
generated in a couple of days on a single 2.8 GHz Xeon processor.

Another challenge to using our approach for high-accuracy DFT on large
systems is the computational expense of using the high-accuracy
subsystem functionals, as this would require performing DFT
calculations in a large basis set. One interesting possibility for
mitigating this expense is to parametrize subsystem functionals that
predict both $^1D$ and $^2\Delta$ in a large basis set as a function
of $^1D$ in a smaller basis. This would enable us to use a database of
high-accuracy, large-basis-set subsystem calculations to correct
small-basis-set DFT on a large system.

We believe that the general approach presented here (see flowchart,
Sec.~\ref{sec:param}) may be useful for modeling several aspects of
electronic structure in addition to intra-subsystem electron
correlation effects. We are currently developing a semiempirical,
subsystem-based treatment of dispersion interactions in DFT that uses
semiempirical functionals to predict a subsystem's polarizability as a
function of its $^1D$. Also, as mentioned above, we are developing a
treatment of core polarization in effective core potentials. This
treatment uses a semiempirical functional to predict the change in
core electron density matrix as a function of the valence density matrix and the
core-electron Hamiltonian.

This work explores a new approach for taking advantage of molecular
similarity in electronic structure theory. The results suggest that it
may be possible to construct accurate semiempirical models by
extracting transferable information from {\em ab initio} data on small
molecules. However, the applicability of the method to larger systems
remains to be explored.

The authors thank Craig J. Gallek for contributions to extensions to
GAMESS for density matrix manipulation. This work was supported by the
National Science Foundation. BGJ thanks the NSF for additional
support.

\clearpage
\newpage
\section{Supporting Information}

\subsection{Sources of error in fixed-geometry \hten\ }

The conclusion that the error in fixed-geometry \hten\ is due to
long-range order is supported by results from the ($C_1=7,\ C_2=8$)
\map\ functional discussed in Sec~\ref{sec:hfourres}. This functional
gave a better description of the fixed-geometry \hfour\ systems than
the functional in Table~\ref{tab:hchaindata}. However, this functional
does {\em not} give a better description of the \hten\ systems: the
$^2\Delta[^1D_{exact}]$ \dEabs\ is 5.01 (11.30) mH, much larger than
the 1.81 (1.99) value in Table~\ref{tab:res2}.

To further confirm that the fixed-geometry error is due to the effects
of long-range order, we parametrized \map\ functionals for a new
fixed-geometry \hten\ system with increased long-range order. This
system was generated as in Sec.~\ref{sec:generate} but with an \htwo\
$\leftrightarrow$ \htwo\ spacing of 1.0 \AA\ rather than 1.6 \AA. Its
\hfour\ subsystems are expected to be even less similar to isolated
\hfour\ molecules. The increased long-range order is seen in an
increased (though still quite small) subsystem decomposition error,
with average (standard deviation) $^2\Delta_{xsub}$ \dEabs\ of {\mbox
1.01 (0.60) mH} vs. the {\mbox 0.02 (0.04) mH} values in
Table~\ref{tab:res2}. This system's \map\ \hfour\ functionals gave
\dEabs\ for \hfour\ comparable to the values in Table~\ref{tab:res1}:
average (standard deviation) values of the $^2\Delta[^1D_{exact}]$
\dEabs\ for the testing-set data are 0.29 (0.43) mH. However, as
expected, the increased long-range order meant that the \map\
functionals parametrized on isolated \hfour\ molecules gave {\em very}
poor results when applied to \hten. The $^2\Delta[^1D_{exact}]$
\dEabs\ was 17.93 (4.89) mH, much larger than the value in
Table~\ref{tab:res2}.

\subsection{Training set choice calculations for \hfour\ \map\ functionals}

The predictions of a semiempirical model should not depend on the
choice of training set data. We tested this by parametrizing several
different \hfour\ \map\ functionals (102 for variable-geometry
systems, 74 for fixed-geometry systems), each with a different
training set choice. Table~\ref{tab:setchoice} presents the average
and standard deviation, taken across the training set choices, of the
average \dEabs\ values of each data set.

To clarify how the results in Table~\ref{tab:setchoice} were obtained,
let $S_{choice}$ denote the set of $N_{choice}$ different choices of
training set tested, where $N_{choice}$ equals 102 and 74 for
variable- and fixed-geometry systems. The training set choices in
$S_{choice}$ are indexed by $x$.  Let $S_{train}^{x}$ denote the set
of \hfour\ molecules in training set $x$, where each molecule is
indexed by $i_x$. Each $S_{train}^{x}$ contains 500 of the 1000 total
\hfour\ molecules (Table~\ref{tab:hchaindata}), with the remainder in
the test set. Let the absolute correlation energy error \dEabs\ for
$^2\Delta[^1D_{exact}]$ of each data point in training set $x$ be
denoted $\left|\delta E_{corr}\right|(i_x)$, and let $AVE\{\ \}$ and
$STDEV\{\ \}$ denote the operations of calculating the average and
standard deviation of a set of points. The average (standard
deviation) values of the first entry in Table~\ref{tab:setchoice} (row
1, column 2), denoted ``A'' and ``B'', are obtained as
\begin{eqnarray}
\label{eq:setchoice}
A = AVE\{\ AVE\{\left|\delta E_{corr}\right|(i_x), i_x \in S_{train}^{x} \},x\in S_{choice} \} \\ 
B = STDEV\{\ AVE\{\left|\delta E_{corr}\right|(i_x),i_x \in S_{train}^{x} \}, x\in S_{choice} \} \nonumber
\end{eqnarray}

The results in Table~\ref{tab:setchoice} verify that the \map\
functional predictions do not depend very much on the training set
choice.

When parametrizing a model, it is useful to test models that were
parametrized with an incorrect input:output relation in the training
set. If the model is implemented correctly, and is modeling a real
physical relationship, scrambling the data should degrade the
results. Table~\ref{tab:setchoice} includes results from a \map\
functional where the $^1D$ from each molecule in the training set is
paired randomly with the $^2\Delta$ of a different molecule. This
scrambles the input:output relation of the training data, and is
denoted $^2\Delta[^1D_{exact}] (scr)$. As expected, this functional is
no better (and sometimes worse) than $^2\Delta[^1D_{avg}]$, which uses
a single choice of $^2\Delta$ for all data points.

\clearpage
\newpage

\bibliography{refs,footnotes}

\begin{thebibliography}{63}
\expandafter\ifx\csname natexlab\endcsname\relax\def\natexlab#1{#1}\fi
\expandafter\ifx\csname bibnamefont\endcsname\relax
  \def\bibnamefont#1{#1}\fi
\expandafter\ifx\csname bibfnamefont\endcsname\relax
  \def\bibfnamefont#1{#1}\fi
\expandafter\ifx\csname citenamefont\endcsname\relax
  \def\citenamefont#1{#1}\fi
\expandafter\ifx\csname url\endcsname\relax
  \def\url#1{\texttt{#1}}\fi
\expandafter\ifx\csname urlprefix\endcsname\relax\def\urlprefix{URL }\fi
\providecommand{\bibinfo}[2]{#2}
\providecommand{\eprint}[2][]{\url{#2}}

\bibitem[{not({\natexlab{a}})}]{note1}
\bibinfo{note}{Full-CI is exact within a given basis set}.

\bibitem[{\citenamefont{Kohn}(1996)}]{kohn_1rdm}
\bibinfo{author}{\bibfnamefont{W.}~\bibnamefont{Kohn}}, \bibinfo{journal}{Phys.
  Rev. Lett.} \textbf{\bibinfo{volume}{76}}, \bibinfo{pages}{3168}
  (\bibinfo{year}{1996}).

\bibitem[{\citenamefont{Goedecker}(1999)}]{goedecker}
\bibinfo{author}{\bibfnamefont{S.}~\bibnamefont{Goedecker}},
  \bibinfo{journal}{Rev. Mod. Phys.} \textbf{\bibinfo{volume}{71}},
  \bibinfo{pages}{1085} (\bibinfo{year}{1999}).

\bibitem[{\citenamefont{Greengard and V.Rokhlin}(1987)}]{fmm}
\bibinfo{author}{\bibfnamefont{L.}~\bibnamefont{Greengard}} \bibnamefont{and}
  \bibinfo{author}{\bibnamefont{V.Rokhlin}}, \bibinfo{journal}{J. Comp. Phys.}
  \textbf{\bibinfo{volume}{73}}, \bibinfo{pages}{325} (\bibinfo{year}{1987}).

\bibitem[{\citenamefont{White et~al.}(1994)\citenamefont{White, Johnson, Gill,
  and Head-Gordon}}]{fmm2}
\bibinfo{author}{\bibfnamefont{C.~A.} \bibnamefont{White}},
  \bibinfo{author}{\bibfnamefont{B.~G.} \bibnamefont{Johnson}},
  \bibinfo{author}{\bibfnamefont{P.~M.~W.} \bibnamefont{Gill}},
  \bibnamefont{and}
  \bibinfo{author}{\bibfnamefont{M.}~\bibnamefont{Head-Gordon}},
  \bibinfo{journal}{Chem. Phys. Lett.} \textbf{\bibinfo{volume}{230}},
  \bibinfo{pages}{8} (\bibinfo{year}{1994}).

\bibitem[{\citenamefont{White et~al.}(1996)\citenamefont{White, Johnson, Gill,
  and Head-Gordon}}]{fmm3}
\bibinfo{author}{\bibfnamefont{C.~A.} \bibnamefont{White}},
  \bibinfo{author}{\bibfnamefont{B.~G.} \bibnamefont{Johnson}},
  \bibinfo{author}{\bibfnamefont{P.~M.~W.} \bibnamefont{Gill}},
  \bibnamefont{and}
  \bibinfo{author}{\bibfnamefont{M.}~\bibnamefont{Head-Gordon}},
  \bibinfo{journal}{Chem. Phys. Lett.} \textbf{\bibinfo{volume}{253}},
  \bibinfo{pages}{268} (\bibinfo{year}{1996}).

\bibitem[{\citenamefont{Yang}(1991)}]{dac_dft}
\bibinfo{author}{\bibfnamefont{W.}~\bibnamefont{Yang}}, \bibinfo{journal}{Phys.
  Rev. Lett.} \textbf{\bibinfo{volume}{66}}, \bibinfo{pages}{1438}
  (\bibinfo{year}{1991}).

\bibitem[{\citenamefont{Bowler et~al.}(2002)\citenamefont{Bowler, Miyazaki, and
  Gillian}}]{conquest}
\bibinfo{author}{\bibfnamefont{D.~R.} \bibnamefont{Bowler}},
  \bibinfo{author}{\bibfnamefont{T.}~\bibnamefont{Miyazaki}}, \bibnamefont{and}
  \bibinfo{author}{\bibfnamefont{M.~J.} \bibnamefont{Gillian}},
  \bibinfo{journal}{Journal of Physics: Condensed Matter}
  \textbf{\bibinfo{volume}{14}} (\bibinfo{year}{2002}).

\bibitem[{\citenamefont{Zhang and Zhang}(2003)}]{mfcc}
\bibinfo{author}{\bibfnamefont{D.~W.} \bibnamefont{Zhang}} \bibnamefont{and}
  \bibinfo{author}{\bibfnamefont{J.~Z.~H.} \bibnamefont{Zhang}},
  \bibinfo{journal}{J. Chem. Phys.} \textbf{\bibinfo{volume}{119}},
  \bibinfo{pages}{3599} (\bibinfo{year}{2003}).

\bibitem[{\citenamefont{Pulay}(1983)}]{ps1}
\bibinfo{author}{\bibfnamefont{P.}~\bibnamefont{Pulay}},
  \bibinfo{journal}{Chem. Phys. Lett.} \textbf{\bibinfo{volume}{100}},
  \bibinfo{pages}{151} (\bibinfo{year}{1983}).

\bibitem[{\citenamefont{Saebo and Pulay}(1993)}]{pulay1}
\bibinfo{author}{\bibfnamefont{S.}~\bibnamefont{Saebo}} \bibnamefont{and}
  \bibinfo{author}{\bibfnamefont{P.}~\bibnamefont{Pulay}},
  \bibinfo{journal}{Ann. Rev. Phys. Chem.} \textbf{\bibinfo{volume}{44}},
  \bibinfo{pages}{213} (\bibinfo{year}{1993}).

\bibitem[{\citenamefont{Hampel and Werner}(1996)}]{werner1}
\bibinfo{author}{\bibfnamefont{C.}~\bibnamefont{Hampel}} \bibnamefont{and}
  \bibinfo{author}{\bibfnamefont{H.-J.} \bibnamefont{Werner}},
  \bibinfo{journal}{J. Chem. Phys.} \textbf{\bibinfo{volume}{104}},
  \bibinfo{pages}{6286} (\bibinfo{year}{1996}).

\bibitem[{\citenamefont{P.E.Maslen and Head-Gordon}(1998)}]{mhg1}
\bibinfo{author}{\bibnamefont{P.E.Maslen}} \bibnamefont{and}
  \bibinfo{author}{\bibfnamefont{M.}~\bibnamefont{Head-Gordon}},
  \bibinfo{journal}{Chem. Phys. Lett.} \textbf{\bibinfo{volume}{283}},
  \bibinfo{pages}{102} (\bibinfo{year}{1998}).

\bibitem[{\citenamefont{Scuseria and Ayala}(1999)}]{as_ccd}
\bibinfo{author}{\bibfnamefont{G.~E.} \bibnamefont{Scuseria}} \bibnamefont{and}
  \bibinfo{author}{\bibfnamefont{P.~Y.} \bibnamefont{Ayala}},
  \bibinfo{journal}{J. Chem. Phys.} \textbf{\bibinfo{volume}{111}},
  \bibinfo{pages}{8330} (\bibinfo{year}{1999}).

\bibitem[{\citenamefont{Sch\"utz and Werner}(2001)}]{werner3}
\bibinfo{author}{\bibfnamefont{M.}~\bibnamefont{Sch\"utz}} \bibnamefont{and}
  \bibinfo{author}{\bibfnamefont{H.-J.} \bibnamefont{Werner}},
  \bibinfo{journal}{J. Chem. Phys.} \textbf{\bibinfo{volume}{114}},
  \bibinfo{pages}{661} (\bibinfo{year}{2001}).

\bibitem[{\citenamefont{Van~Voorhis and Head-Gordon}(2001)}]{mhg3}
\bibinfo{author}{\bibfnamefont{T.}~\bibnamefont{Van~Voorhis}} \bibnamefont{and}
  \bibinfo{author}{\bibfnamefont{M.}~\bibnamefont{Head-Gordon}},
  \bibinfo{journal}{J. Chem. Phys.} \textbf{\bibinfo{volume}{115}},
  \bibinfo{pages}{7814} (\bibinfo{year}{2001}).

\bibitem[{\citenamefont{Li et~al.}(2002)\citenamefont{Li, Ma, and Jiang}}]{lmj}
\bibinfo{author}{\bibfnamefont{S.}~\bibnamefont{Li}},
  \bibinfo{author}{\bibfnamefont{J.}~\bibnamefont{Ma}}, \bibnamefont{and}
  \bibinfo{author}{\bibfnamefont{Y.}~\bibnamefont{Jiang}}, \bibinfo{journal}{J.
  Comp. Chem.} \textbf{\bibinfo{volume}{23}}, \bibinfo{pages}{237}
  (\bibinfo{year}{2002}).

\bibitem[{\citenamefont{Flocke and Bartlett}(2003)}]{bart}
\bibinfo{author}{\bibfnamefont{N.}~\bibnamefont{Flocke}} \bibnamefont{and}
  \bibinfo{author}{\bibfnamefont{R.~J.} \bibnamefont{Bartlett}},
  \bibinfo{journal}{J. Chem. Phys.} \textbf{\bibinfo{volume}{118}}
  (\bibinfo{year}{2003}).

\bibitem[{\citenamefont{Machida}(1999)}]{mm}
\bibinfo{author}{\bibfnamefont{K.}~\bibnamefont{Machida}},
  \emph{\bibinfo{title}{Principles of molecular mechanics}}
  (\bibinfo{publisher}{Wiley}, \bibinfo{address}{New York},
  \bibinfo{year}{1999}).

\bibitem[{\citenamefont{Ridley and Zerner}(1973)}]{indo}
\bibinfo{author}{\bibfnamefont{J.}~\bibnamefont{Ridley}} \bibnamefont{and}
  \bibinfo{author}{\bibfnamefont{M.}~\bibnamefont{Zerner}},
  \bibinfo{journal}{Theoretica Chemica Acta} \textbf{\bibinfo{volume}{32}},
  \bibinfo{pages}{111} (\bibinfo{year}{1973}).

\bibitem[{\citenamefont{Zoebisch et~al.}(1985)\citenamefont{Zoebisch, Healey,
  Stewart, and Dewar}}]{am1}
\bibinfo{author}{\bibfnamefont{E.~J.} \bibnamefont{Zoebisch}},
  \bibinfo{author}{\bibfnamefont{E.~F.} \bibnamefont{Healey}},
  \bibinfo{author}{\bibfnamefont{J.~J.~P.} \bibnamefont{Stewart}},
  \bibnamefont{and} \bibinfo{author}{\bibfnamefont{M.~J.~S.}
  \bibnamefont{Dewar}}, \bibinfo{journal}{JACS} \textbf{\bibinfo{volume}{107}},
  \bibinfo{pages}{3902} (\bibinfo{year}{1985}).

\bibitem[{\citenamefont{Warshel and Karplus}(1972)}]{qmmm1a}
\bibinfo{author}{\bibfnamefont{A.}~\bibnamefont{Warshel}} \bibnamefont{and}
  \bibinfo{author}{\bibfnamefont{M.~J.} \bibnamefont{Karplus}},
  \bibinfo{journal}{JACS} \textbf{\bibinfo{volume}{94}}, \bibinfo{pages}{5612}
  (\bibinfo{year}{1972}).

\bibitem[{\citenamefont{Maseras and Morokuma}(1995)}]{qmmm2}
\bibinfo{author}{\bibfnamefont{F.}~\bibnamefont{Maseras}} \bibnamefont{and}
  \bibinfo{author}{\bibfnamefont{K.}~\bibnamefont{Morokuma}},
  \bibinfo{journal}{J. Comp. Chem.} \textbf{\bibinfo{volume}{16}},
  \bibinfo{pages}{1170} (\bibinfo{year}{1995}).

\bibitem[{\citenamefont{Ercolessi and Adams}(1994)}]{fromab_4}
\bibinfo{author}{\bibfnamefont{F.}~\bibnamefont{Ercolessi}} \bibnamefont{and}
  \bibinfo{author}{\bibfnamefont{J.}~\bibnamefont{Adams}},
  \bibinfo{journal}{Europhysics Letters} \textbf{\bibinfo{volume}{26}},
  \bibinfo{pages}{583} (\bibinfo{year}{1994}).

\bibitem[{\citenamefont{Mehl and Papaconstantopoulos}(1996)}]{fromab_1}
\bibinfo{author}{\bibfnamefont{M.~J.} \bibnamefont{Mehl}} \bibnamefont{and}
  \bibinfo{author}{\bibfnamefont{D.~A.} \bibnamefont{Papaconstantopoulos}},
  \bibinfo{journal}{Phys. Rev. B} \textbf{\bibinfo{volume}{54}},
  \bibinfo{pages}{4519} (\bibinfo{year}{1996}).

\bibitem[{\citenamefont{Tabacchi et~al.}(2002)\citenamefont{Tabacchi, Mundy,
  Hutter, and Parrinello}}]{fromab_2}
\bibinfo{author}{\bibfnamefont{G.}~\bibnamefont{Tabacchi}},
  \bibinfo{author}{\bibfnamefont{C.~J.} \bibnamefont{Mundy}},
  \bibinfo{author}{\bibfnamefont{J.}~\bibnamefont{Hutter}}, \bibnamefont{and}
  \bibinfo{author}{\bibfnamefont{M.}~\bibnamefont{Parrinello}},
  \bibinfo{journal}{J. Chem. Phys.} \textbf{\bibinfo{volume}{117}},
  \bibinfo{pages}{1416} (\bibinfo{year}{2002}).

\bibitem[{\citenamefont{Tangney and Scandolo}(2002)}]{fromab_3}
\bibinfo{author}{\bibfnamefont{P.}~\bibnamefont{Tangney}} \bibnamefont{and}
  \bibinfo{author}{\bibfnamefont{S.}~\bibnamefont{Scandolo}},
  \bibinfo{journal}{J. Chem. Phys.} \textbf{\bibinfo{volume}{117}},
  \bibinfo{pages}{8898} (\bibinfo{year}{2002}).

\bibitem[{not({\natexlab{b}})}]{note2}
\bibinfo{note}{In a non-orthogonal basis like those used here, the real-space
  electron density $^1D({\mathbf r})$ and the real-space density matrix
  $^1D({\mathbf r},{\mathbf r^{'}})$ are obtained from $^1D(a,b)$ as
  $^1D({\mathbf r})=\sum_{ab}\phi_a^*({\mathbf r})\phi_b({\mathbf r})\
  ^1D(a,b)$ and $^1D({\mathbf r},{\mathbf r^{'}})=\sum_{ab}\phi_a^*({\mathbf
  r})\phi_b({\mathbf r^{'}})\ ^1D(a,b)$.}

\bibitem[{\citenamefont{Parr and Yang}(1989)}]{parr}
\bibinfo{author}{\bibfnamefont{R.~G.} \bibnamefont{Parr}} \bibnamefont{and}
  \bibinfo{author}{\bibfnamefont{W.}~\bibnamefont{Yang}},
  \emph{\bibinfo{title}{Density-Functional Theory of Atoms and Molecules}}
  (\bibinfo{publisher}{Oxford University Press}, \bibinfo{address}{New York},
  \bibinfo{year}{1989}).

\bibitem[{\citenamefont{Mazziotti}(1999{\natexlab{a}})}]{maz2}
\bibinfo{author}{\bibfnamefont{D.~A.} \bibnamefont{Mazziotti}},
  \bibinfo{journal}{Phys. Rev. A} \textbf{\bibinfo{volume}{60}},
  \bibinfo{pages}{4396} (\bibinfo{year}{1999}{\natexlab{a}}).

\bibitem[{\citenamefont{Mazziotti}(1999{\natexlab{b}})}]{maz_c2}
\bibinfo{author}{\bibfnamefont{D.~A.} \bibnamefont{Mazziotti}},
  \bibinfo{journal}{Phys. Rev. A} \textbf{\bibinfo{volume}{60}},
  \bibinfo{pages}{3618} (\bibinfo{year}{1999}{\natexlab{b}}).

\bibitem[{\citenamefont{Hohenburg and Kohn}(1964)}]{hk}
\bibinfo{author}{\bibfnamefont{P.}~\bibnamefont{Hohenburg}} \bibnamefont{and}
  \bibinfo{author}{\bibfnamefont{W.}~\bibnamefont{Kohn}},
  \bibinfo{journal}{Phys. Rev.} \textbf{\bibinfo{volume}{136}},
  \bibinfo{pages}{b864} (\bibinfo{year}{1964}).

\bibitem[{\citenamefont{Kohn and Sham}(1965)}]{ks}
\bibinfo{author}{\bibfnamefont{W.}~\bibnamefont{Kohn}} \bibnamefont{and}
  \bibinfo{author}{\bibfnamefont{L.~J.} \bibnamefont{Sham}},
  \bibinfo{journal}{Phys. Rev.} \textbf{\bibinfo{volume}{140}},
  \bibinfo{pages}{A1133} (\bibinfo{year}{1965}).

\bibitem[{\citenamefont{Burke and Perdew}(1995)}]{vxc_constraints}
\bibinfo{author}{\bibfnamefont{K.}~\bibnamefont{Burke}} \bibnamefont{and}
  \bibinfo{author}{\bibfnamefont{J.~P.} \bibnamefont{Perdew}},
  \bibinfo{journal}{International Journal of Quantum Chemistry}
  \textbf{\bibinfo{volume}{56}}, \bibinfo{pages}{199} (\bibinfo{year}{1995}).

\bibitem[{\citenamefont{Gunnarson et~al.}(1979)\citenamefont{Gunnarson, Jonson,
  and Lundqvist}}]{xchole_nucleii}
\bibinfo{author}{\bibfnamefont{O.}~\bibnamefont{Gunnarson}},
  \bibinfo{author}{\bibfnamefont{M.}~\bibnamefont{Jonson}}, \bibnamefont{and}
  \bibinfo{author}{\bibfnamefont{B.~I.} \bibnamefont{Lundqvist}},
  \bibinfo{journal}{Phys. Rev. B} \textbf{\bibinfo{volume}{20}},
  \bibinfo{pages}{3136} (\bibinfo{year}{1979}).

\bibitem[{\citenamefont{Alonso and Girfalco}(1978)}]{xhole}
\bibinfo{author}{\bibfnamefont{J.~A.} \bibnamefont{Alonso}} \bibnamefont{and}
  \bibinfo{author}{\bibfnamefont{L.~A.} \bibnamefont{Girfalco}},
  \bibinfo{journal}{Phys. Rev. B} \textbf{\bibinfo{volume}{17}},
  \bibinfo{pages}{3735} (\bibinfo{year}{1978}).

\bibitem[{\citenamefont{Dixon and Merz}(1996)}]{dac_semM}
\bibinfo{author}{\bibfnamefont{S.~L.} \bibnamefont{Dixon}} \bibnamefont{and}
  \bibinfo{author}{\bibfnamefont{K.~M.} \bibnamefont{Merz},
  \bibfnamefont{Jr.}}, \bibinfo{journal}{J. Chem. Phys.}
  \textbf{\bibinfo{volume}{104}}, \bibinfo{pages}{6643} (\bibinfo{year}{1996}).

\bibitem[{\citenamefont{Lee et~al.}(1996)\citenamefont{Lee, York, and
  Yang}}]{dac_semY}
\bibinfo{author}{\bibfnamefont{T.-S.} \bibnamefont{Lee}},
  \bibinfo{author}{\bibfnamefont{D.~M.} \bibnamefont{York}}, \bibnamefont{and}
  \bibinfo{author}{\bibfnamefont{W.}~\bibnamefont{Yang}}, \bibinfo{journal}{J.
  Chem. Phys.} \textbf{\bibinfo{volume}{105}}, \bibinfo{pages}{2744}
  (\bibinfo{year}{1996}).

\bibitem[{\citenamefont{Kudin and Scuseria}(2000)}]{gauss_dac_dft}
\bibinfo{author}{\bibfnamefont{K.~N.} \bibnamefont{Kudin}} \bibnamefont{and}
  \bibinfo{author}{\bibfnamefont{G.~E.} \bibnamefont{Scuseria}},
  \bibinfo{journal}{Phys. Rev. B} \textbf{\bibinfo{volume}{61}},
  \bibinfo{pages}{16440} (\bibinfo{year}{2000}).

\bibitem[{\citenamefont{amd K.~Tada et~al.}(2001)\citenamefont{amd K.~Tada,
  Watanabe, Fujita, and Watanabe}}]{dac_dft_bias}
\bibinfo{author}{\bibfnamefont{N.~N.} \bibnamefont{amd K.~Tada}},
  \bibinfo{author}{\bibfnamefont{S.}~\bibnamefont{Watanabe}},
  \bibinfo{author}{\bibfnamefont{H.}~\bibnamefont{Fujita}}, \bibnamefont{and}
  \bibinfo{author}{\bibfnamefont{K.}~\bibnamefont{Watanabe}},
  \bibinfo{journal}{Phys. Rev. Lett.} \textbf{\bibinfo{volume}{86}},
  \bibinfo{pages}{540} (\bibinfo{year}{2001}).

\bibitem[{\citenamefont{Janesko and Yaron}(2003)}]{lrdm}
\bibinfo{author}{\bibfnamefont{B.~G.} \bibnamefont{Janesko}} \bibnamefont{and}
  \bibinfo{author}{\bibfnamefont{D.}~\bibnamefont{Yaron}}, \bibinfo{journal}{J.
  Chem. Phys.} \textbf{\bibinfo{volume}{119}}, \bibinfo{pages}{1320}
  (\bibinfo{year}{2003}).

\bibitem[{\citenamefont{Kohn et~al.}(1998)\citenamefont{Kohn, Meir, and
  Makarov}}]{dft_vdw1}
\bibinfo{author}{\bibfnamefont{W.}~\bibnamefont{Kohn}},
  \bibinfo{author}{\bibfnamefont{Y.}~\bibnamefont{Meir}}, \bibnamefont{and}
  \bibinfo{author}{\bibfnamefont{D.~E.} \bibnamefont{Makarov}},
  \bibinfo{journal}{Phys. Rev. Lett.} \textbf{\bibinfo{volume}{80}},
  \bibinfo{pages}{4153} (\bibinfo{year}{1998}).

\bibitem[{\citenamefont{Andersson et~al.}(1996)\citenamefont{Andersson,
  Langreth, and Lundqvist}}]{dft_vdw2}
\bibinfo{author}{\bibfnamefont{Y.}~\bibnamefont{Andersson}},
  \bibinfo{author}{\bibfnamefont{D.~C.} \bibnamefont{Langreth}},
  \bibnamefont{and} \bibinfo{author}{\bibfnamefont{B.~I.}
  \bibnamefont{Lundqvist}}, \bibinfo{journal}{Phys. Rev. Lett.}
  \textbf{\bibinfo{volume}{76}}, \bibinfo{pages}{102} (\bibinfo{year}{1996}).

\bibitem[{\citenamefont{Jones and Gunnarson}(1989)}]{ldarev}
\bibinfo{author}{\bibfnamefont{R.~O.} \bibnamefont{Jones}} \bibnamefont{and}
  \bibinfo{author}{\bibfnamefont{O.}~\bibnamefont{Gunnarson}},
  \bibinfo{journal}{Rev. Mod. Phys.} \textbf{\bibinfo{volume}{61}},
  \bibinfo{pages}{689} (\bibinfo{year}{1989}).

\bibitem[{\citenamefont{Langreth and Mehl}(1983)}]{gga1}
\bibinfo{author}{\bibfnamefont{D.~C.} \bibnamefont{Langreth}} \bibnamefont{and}
  \bibinfo{author}{\bibfnamefont{M.~J.} \bibnamefont{Mehl}},
  \bibinfo{journal}{Phys. Rev. B} \textbf{\bibinfo{volume}{28}},
  \bibinfo{pages}{1809} (\bibinfo{year}{1983}).

\bibitem[{\citenamefont{Becke}(1993)}]{b3lyp}
\bibinfo{author}{\bibfnamefont{A.~D.} \bibnamefont{Becke}},
  \bibinfo{journal}{J. Chem. Phys.} \textbf{\bibinfo{volume}{98}},
  \bibinfo{pages}{5648} (\bibinfo{year}{1993}).

\bibitem[{\citenamefont{van~de Wall and Ceder}(1999)}]{overbind}
\bibinfo{author}{\bibfnamefont{A.}~\bibnamefont{van~de Wall}} \bibnamefont{and}
  \bibinfo{author}{\bibfnamefont{G.}~\bibnamefont{Ceder}},
  \bibinfo{journal}{Phys. Rev. B} \textbf{\bibinfo{volume}{59}},
  \bibinfo{pages}{14992} (\bibinfo{year}{1999}).

\bibitem[{\citenamefont{Kohn and Mattsson}(1998)}]{edge}
\bibinfo{author}{\bibfnamefont{W.}~\bibnamefont{Kohn}} \bibnamefont{and}
  \bibinfo{author}{\bibfnamefont{A.~E.} \bibnamefont{Mattsson}},
  \bibinfo{journal}{Phys. Rev. Lett.} \textbf{\bibinfo{volume}{81}},
  \bibinfo{pages}{3487} (\bibinfo{year}{1998}).

\bibitem[{\citenamefont{Armiento and Mattsson}(2002)}]{vxsub}
\bibinfo{author}{\bibfnamefont{R.}~\bibnamefont{Armiento}} \bibnamefont{and}
  \bibinfo{author}{\bibfnamefont{A.~E.} \bibnamefont{Mattsson}},
  \bibinfo{journal}{Phys. Rev. B} \textbf{\bibinfo{volume}{66}},
  \bibinfo{pages}{165117} (\bibinfo{year}{2002}).

\bibitem[{\citenamefont{Tozer et~al.}(1996)\citenamefont{Tozer, Ingamells, and
  Handy}}]{nnVxc}
\bibinfo{author}{\bibfnamefont{D.~J.} \bibnamefont{Tozer}},
  \bibinfo{author}{\bibfnamefont{V.~E.} \bibnamefont{Ingamells}},
  \bibnamefont{and} \bibinfo{author}{\bibfnamefont{N.~C.} \bibnamefont{Handy}},
  \bibinfo{journal}{J. Chem. Phys.} \textbf{\bibinfo{volume}{105}},
  \bibinfo{pages}{9200} (\bibinfo{year}{1996}).

\bibitem[{\citenamefont{Zhao et~al.}(1994)\citenamefont{Zhao, Morrison, and
  Parr}}]{abVxc_1}
\bibinfo{author}{\bibfnamefont{Q.}~\bibnamefont{Zhao}},
  \bibinfo{author}{\bibfnamefont{R.~C.} \bibnamefont{Morrison}},
  \bibnamefont{and} \bibinfo{author}{\bibfnamefont{R.~G.} \bibnamefont{Parr}},
  \bibinfo{journal}{Phys. Rev. A} \textbf{\bibinfo{volume}{50}},
  \bibinfo{pages}{2138} (\bibinfo{year}{1994}).

\bibitem[{\citenamefont{Gritsenko et~al.}(1995)\citenamefont{Gritsenko, van
  Leeuwen, and Baerends}}]{abVxc_2}
\bibinfo{author}{\bibfnamefont{O.~V.} \bibnamefont{Gritsenko}},
  \bibinfo{author}{\bibfnamefont{R.}~\bibnamefont{van Leeuwen}},
  \bibnamefont{and} \bibinfo{author}{\bibfnamefont{E.~J.}
  \bibnamefont{Baerends}}, \bibinfo{journal}{Phys. Rev. A}
  \textbf{\bibinfo{volume}{52}}, \bibinfo{pages}{1870} (\bibinfo{year}{1995}).

\bibitem[{\citenamefont{Cherkassky and Mulier}(1998)}]{databook}
\bibinfo{author}{\bibfnamefont{V.}~\bibnamefont{Cherkassky}} \bibnamefont{and}
  \bibinfo{author}{\bibfnamefont{F.}~\bibnamefont{Mulier}},
  \emph{\bibinfo{title}{Learning from Data: Concepts, Theory, and Methods}}
  (\bibinfo{publisher}{Wiley-Interscience}, \bibinfo{year}{1998}).

\bibitem[{\citenamefont{Rauhut et~al.}(1995)\citenamefont{Rauhut, Boughton, and
  Pulay}}]{nnEcorr}
\bibinfo{author}{\bibfnamefont{G.}~\bibnamefont{Rauhut}},
  \bibinfo{author}{\bibfnamefont{J.~W.} \bibnamefont{Boughton}},
  \bibnamefont{and} \bibinfo{author}{\bibfnamefont{P.}~\bibnamefont{Pulay}},
  \bibinfo{journal}{J. Chem. Phys.} \textbf{\bibinfo{volume}{103}},
  \bibinfo{pages}{5662} (\bibinfo{year}{1995}).

\bibitem[{\citenamefont{Dolg}(2000)}]{ecp}
\bibinfo{author}{\bibfnamefont{M.}~\bibnamefont{Dolg}}, in
  \emph{\bibinfo{booktitle}{Modern Methods and Algorithms of Quantum Chemistry,
  NIC Series 1}}, edited by
  \bibinfo{editor}{\bibfnamefont{J.}~\bibnamefont{Grotendorst}}
  (\bibinfo{publisher}{John Neumann Institute for Computing},
  \bibinfo{year}{2000}), pp. \bibinfo{pages}{479 -- 508}.

\bibitem[{not({\natexlab{c}})}]{note3}
\bibinfo{note}{{The number of principle components required to describe an
  M-orbital $^1D$ should scale asymptotically as ${\cal O}(M)$, but our
  subsystems are designed to be too small for this nearsighted assumption.}}

\bibitem[{not({\natexlab{d}})}]{note4}
\bibinfo{note}{{The dimensionally-reduced subsystem blocks of $^2\Delta_{PCA}$
  are recombined using LRDM}}.

\bibitem[{\citenamefont{Schmidt et~al.}(1993)}]{GAMESS}
\bibinfo{author}{\bibfnamefont{M.~W.} \bibnamefont{Schmidt}}
  \bibnamefont{et~al.}, \bibinfo{journal}{J. Comput. Chem.}
  \textbf{\bibinfo{volume}{14}}, \bibinfo{pages}{1347} (\bibinfo{year}{1993}).

\bibitem[{not({\natexlab{e}})}]{note5}
\bibinfo{note}{{LRDM on minimal-basis \hten\ does not discard any $^2\Delta$
  information from the subsystem edges, as the subsystems are very small.}}

\bibitem[{not({\natexlab{f}})}]{note6}
\bibinfo{note}{{Except for the central aldehyde carbon, the Cartesian
  coordinates of each atom in \ald{R} were perturbed by a random variable
  $\delta$ where $\left|\delta\right|\leq 0.1 \AA$}}.

\bibitem[{not({\natexlab{g}})}]{note7}
\bibinfo{note}{{10 random fractional charges placed in a cube, 8.0 \AA\ to a
  side, centered on the aldehyde carbon atom with $\geq 1.2 \AA$ charge-atom
  separation}}.

\bibitem[{\citenamefont{Pople et~al.}(1979)\citenamefont{Pople, Krishnan,
  Schlegel, and Binkley}}]{relax1}
\bibinfo{author}{\bibfnamefont{J.~A.} \bibnamefont{Pople}},
  \bibinfo{author}{\bibfnamefont{R.}~\bibnamefont{Krishnan}},
  \bibinfo{author}{\bibfnamefont{H.~B.} \bibnamefont{Schlegel}},
  \bibnamefont{and} \bibinfo{author}{\bibfnamefont{J.~S.}
  \bibnamefont{Binkley}}, \bibinfo{journal}{International Journal of Quantum
  Chemistry: Quantum Chemistry Symposium} \textbf{\bibinfo{volume}{13}}
  (\bibinfo{year}{1979}).

\bibitem[{\citenamefont{Dunning}(1989)}]{ccbasis}
\bibinfo{author}{\bibfnamefont{T.~H.} \bibnamefont{Dunning},
  \bibfnamefont{Jr.}}, \bibinfo{journal}{J. Chem. Phys.}
  \textbf{\bibinfo{volume}{90}}, \bibinfo{pages}{1007} (\bibinfo{year}{1989}).

\end{thebibliography}

\clearpage
\newpage
\begin{table}
\begin{tabular}[c]{l@{\hspace{0.2in}} c c c }
Approximation & $^2\Delta_{xsub}$ & $^2\Delta_{PCA}$ & $^2\Delta[^1D_{exact}]$ \\ 
\hline
Subsystem decomposition & Yes & Yes & Yes \\ 
Dimensional reduction & No & Yes & Yes \\ 
Prediction from $^1D$ & No & No & Yes \\ 
\end{tabular}
\caption{\label{tab:rho2types} Types of connected pair density matrix
$^2\Delta$ obtained in the results, and the approximations associated
with each.}
\end{table}

\clearpage
\newpage

\begin{table}
\begin{tabular}[c]{l@{\hspace{0.12in}} c c @{\extracolsep{0.1in}} c c }
 & \multicolumn{2}{c}{Variable geometry} & \multicolumn{2}{c}{Fixed geometry} \\ 
\cline{2-3}\cline{4-5}
 & \hfour\ & \hten\ & \hfour\ & \hten\ \\ 
\hline
$N_{dat}$ & 1000 & 93 & 1000 & 99 \\ 
Point charges & 4 & 10 & 4 & 10 \\
\hline
\htwo\ bonds & \multicolumn{2}{c}{0.5 $\leftrightarrow$ 1.0 \AA} & \multicolumn{2}{c}{0.7 \AA} \\ 
\htwo\ $\leftrightarrow$ \htwo\  &  \multicolumn{2}{c}{0.9 $\leftrightarrow$ 3.0 \AA} & \multicolumn{2}{c}{1.6 \AA} \\ 
$C_1$  & \multicolumn{2}{c}{6} &  \multicolumn{2}{c}{4} \\ 
$C_2$ & \multicolumn{2}{c}{5} &  \multicolumn{2}{c}{5} \\ 
\end{tabular}
\caption{\label{tab:hchaindata} Details of the four data sets for
linear dimerized hydrogen chains. $N_{dat}$ is the total number of
molecules in the data set. $C_1$ and $C_2$ are the number of $^1D$
and $^2\Delta$ principal components used in the \map\ functionals
(Eq.~\ref{eq:twodeltapca}). }
\end{table}

\clearpage
\newpage

\begin{table}
\begin{tabular}[c]{l c c c c }
Prediction & V train & V test & F train & F test  \\
\hline
$^2\Delta[^1D_{exact}]$ & 1.95 (1.61)  & 1.91 (1.85)  & 0.37 (0.51)  & 0.46 (0.96)\\
$^2\Delta_{PCA}$ & 1.29 (1.29)  & 1.27 (1.22)  & 0.11 (0.16)  & 0.13 (0.27)\\
$^2\Delta[^1D_{ave}]$ & 3.78 (3.77)  & 3.66 (3.47)  & 1.53 (2.72)  & 1.49 (2.35)\\
MP2 &\multicolumn{2}{c}{85.83 (18.93) } &\multicolumn{2}{c}{68.52 (2.04) }\\
\end{tabular}
\caption{\label{tab:res1} Absolute $E_{corr}$ error $\left|\delta
E_{corr}\right|$ (mH) 
for training and testing subsets of the variable- and fixed-geometry
\hfour\ subsystems (V and F, respectively). Values are average
(standard deviation) across the entire training or testing set, for a
single choice of training set. MP2 \dEabs\ values are included for
comparison. The average and standard deviation of the correct
$E_{corr}$ values are {\mbox -114.39 (23.43) mH} for the
variable-geometry \hfour\ and {\mbox -93.10 (2.56) mH} for the
fixed-geometry \hfour.  }
\end{table}

\clearpage
\newpage

\begin{table}
\begin{tabular}[c]{ l c c }
Prediction & V & F \\
\hline
$^2\Delta[^1D_{exact}]$ & 3.43 (2.96)  & 1.81 (1.99)  \\
$^2\Delta_{xsub}$ & 0.02 (0.04)  & 0.02 (0.04)  \\
$^2\Delta_{PCA}$ & 3.33 (3.16)  & 1.40 (1.21)  \\
$^2\Delta[^1D_{avg}]$ & 9.21 (6.53)  & 2.55 (5.06)  \\
MP2 & 218.14 (30.06)  & 169.61 (4.33)  \\
\end{tabular}
\caption{\label{tab:res2} Absolute $E_{corr}$ error $\left|\delta
E_{corr}\right|$ (mH) for variable- and fixed-geometry \hten\ (V and F,
respectively). Values are average (standard deviation) across the
entire data set for the choice of training set used in
Table~\ref{tab:res1}. MP2 \dEabs\ values are included for
comparison. }
\end{table}

\clearpage
\newpage

\begin{table}
\begin{tabular}[c]{l l c c }
System & Prediction & V & F \\
\hline
\hfour\ & $^2\Delta[^1D_{HF}]$ & 5.72 (3.91)  & 3.21 (1.02) \\
 & $^2\Delta[^1D_{DFT}]$ & 4.10 (3.34)  & 1.24 (0.66) \\
\hline
\hten\ & $^2\Delta[^1D_{HF}]$ & 15.73 (5.46)  & 9.24 (1.48) \\
 & $^2\Delta[^1D_{DFT}]$ & 12.02 (4.73)  & 4.36 (1.90) \\
\end{tabular}
\caption{\label{tab:dft1} Absolute $E_{corr}$ error $\left|\delta
E_{corr}\right|$ (mH) for DFT and corrected Hartree-Fock calculations
using \map\ correlation energy functionals ($^2\Delta[^1D_{HF}]$ and
$^2\Delta[^1D_{DFT}]$, respectively). Results are presented for
\hfour\ and \hten, variable (V) and fixed (F) geometry hydrogen
chains, average (standard deviation) over the entire data set for a
single training set choice.  }
\end{table}

\clearpage
\newpage

\begin{table}
\begin{tabular}[c]{l@{\hspace{0.12in}} c c c c c c }
Excluded & H & F & OH & CH$_3$ & Cl & OCH$_3$\\
\hline
H & {\bf 2.56} & 1.05 & 1.11 & 1.17 & 1.09 & 1.10\\
F & 1.44 & {\bf 3.24} & 1.19 & 1.42 & 1.31 & 1.23\\
OH & 1.33 & 1.42 & {\bf 1.42} & 1.50 & 1.21 & 1.47\\
CH$_3$ & 1.43 & 1.27 & 1.23 & {\bf 2.35} & 1.32 & 1.30\\
Cl & 1.55 & 1.16 & 1.18 & 1.57 & {\bf 8.10} & 1.34\\
OCH$_3$ & 1.37 & 1.26 & 1.23 & 1.47 & 1.19 & {\bf 1.76} \\
\end{tabular}

\vspace{0.1in}

\begin{tabular}[c]{l@{\hspace{0.12in}} c c c c c c }
Excluded & H & F & OH & CH$_3$ & Cl & OCH$_3$\\
\hline
H & {\bf 6.00} & 5.50 & 4.48 & 4.48 & 6.62 & 5.37\\
F & 6.07 & {\bf 5.44} & 5.09 & 4.80 & 6.47 & 4.88\\
OH & 5.50 & 4.36 & {\bf 4.70} & 4.11 & 8.13 & 5.05\\
CH$_3$ & 6.37 & 5.83 & 4.76 & {\bf 4.69} & 6.88 & 5.52\\
Cl & 5.01 & 4.82 & 4.85 & 4.28 & {\bf 9.09} & 5.09\\
OCH$_3$ & 5.67 & 5.17 & 4.74 & 4.35 & 7.70 & {\bf 5.12} \\
\end{tabular}

\caption{\label{tab:ald1}Absolute \Esub\ errors (mH) for the six
different \ald{R} \map\ functionals.  The rows are the results for
each of the six functionals, where the R group that was excluded from
each functional's training data is listed in the first column. The
columns show the mean absolute \Esub\ error for each of the six kinds
of \ald{R} molecules in the testing set. Extrapolations to the R group
excluded from each functional are shown in boldface. Upper and lower panels are
data for $^2\Delta[^1D_{exact}]$ and $^2\Delta[^1D_{avg}]$. The
extrapolation results for $^2\Delta[^1D_{exact}]$ are plotted in
Fig.~\ref{fig:aldplot2}.}
\end{table}

\clearpage
\newpage

\begin{table}
\begin{tabular}[c]{l c c c}
System & $^2\Delta[^1D_{exact}]$ & $^2\Delta[^1D_{avg}]$ & $^2\Delta[^1D_{exact}] (scr)$ \\ 
\hline
V train & 1.75 (0.15) & 3.70 (0.12) & 6.93 (0.33) \\
V test & 1.83 (0.15) & 3.75 (0.13) & 7.00 (0.39) \\
V \hten\ & 3.12 (0.23) & 10.18 (0.69) & 12.91 (0.74) \\
F train & 0.45 (0.04) & 1.38 (0.11) & 1.43 (0.12) \\
F test & 0.48 (0.03) & 1.36 (0.15) & 1.41 (0.08) \\
F \hten\ & 1.98 (0.17) & 2.53 (0.04) & 3.95 (0.30) \\
\end{tabular}
\caption{\label{tab:setchoice} Absolute $E_{corr}$ errors
$\left|\delta E_{corr}\right|$ (mH) for \hfour\ \map\ functionals, for
multiple choices of training set.  Each entry is the average value of
all molecules in the training or testing data set, average (standard
deviation) over the training set choices (see text for
details). Results are reported for variable- and fixed-geometry
systems (respectively V and F), for \hfour\ training and testing sets
and extrapolation to \hten\ (respectively train, test, and \hten ).}
\end{table}

\clearpage
\newpage

{\bf Figure captions}
\begin{figure}[htb]
\caption{\label{fig:lrdm} Schematic of a nearsightedness-based
divide-and-conquer treatment of electronic structure for a generic
four-element chain. The electronic structure of the three subsystems
(boxed regions) are obtained separately (calculations I-III) and
combined into an approximate electronic structure for the entire
system (``Total''). The calculated electronic structure near the edges
of each subsystem (dotted lines) is incorrect due to short-range edge
effects, and is not used in the final approximate structure.}
\end{figure}
\newline
\newline
\begin{figure}
\caption{\label{fig:res1}Predicted vs. real $E_{corr}$ (mH) for
variable- and fixed-geometry \hfour\ (A and B). The
correlation coefficient R$^2$ between real and predicted $E_{corr}$
are in parentheses. To reduce congestion, the variable-geometry
$^2\Delta_{PCA}$ and $^2\Delta[^1D_{ave}]$ $E_{corr}$ are shifted down
by 30 and {\mbox 60 mH}. MP2 $E_{corr}$ are shifted down by 86 and
{\mbox 67 mH} for the variable- and fixed-geometry results,
respectively.}
\end{figure}
\newline
\newline
\begin{figure}
\caption{\label{fig:res2}Predicted vs. real correlation energies (mH)
for variable- and fixed-geometry \hten\ (A and B),
predicted using \hfour\ functionals.  R$^2$ between real and predicted
$E_{corr}$ are in parentheses. To reduce congestion, the variable-geometry
$^2\Delta_{xsub}$, $^2\Delta_{PCA}$ and $^2\Delta[^1D_{ave}]$ $E_{corr}$ are shifted down
by 60, 60, and {\mbox 120 mH}, respectively. MP2 $E_{corr}$ are shifted down by 216 and
{\mbox 163 mH} for the variable- and fixed-geometry results,
respectively.}
\end{figure}
\newline
\newline
\begin{figure}
\caption{\label{fig:dft2}Predicted vs. real correlation energies (mH)
for DFT and corrected Hartree-Fock calculations using \hfour\ \map\
functionals. Results are presented for variable- and fixed-geometry
\hten\ (A and B). R$^2$ between real and predicted
$E_{corr}$ are in parentheses. }
\end{figure}
\newline
\newline
\begin{figure}
\caption{\label{fig:aldplot2} Extrapolation results. Predicted
vs. real $^2\Delta[^1D_{exact}]$ \Esub\ for the six kinds of \ald{R}
molecules. Each of the \ald{R} data sets is modeled using the \map\
functional that was not trained on data from that R group. The
correlation coefficients R$^2$ between real and predicted \Esub\ are
in parentheses. Absolute \Esub\ errors for the plotted data are the
diagonal (boldface) entries in the upper panel of
Table~\ref{tab:ald1}.}
\end{figure}

\clearpage
\newpage
\includegraphics[width=3.0in]{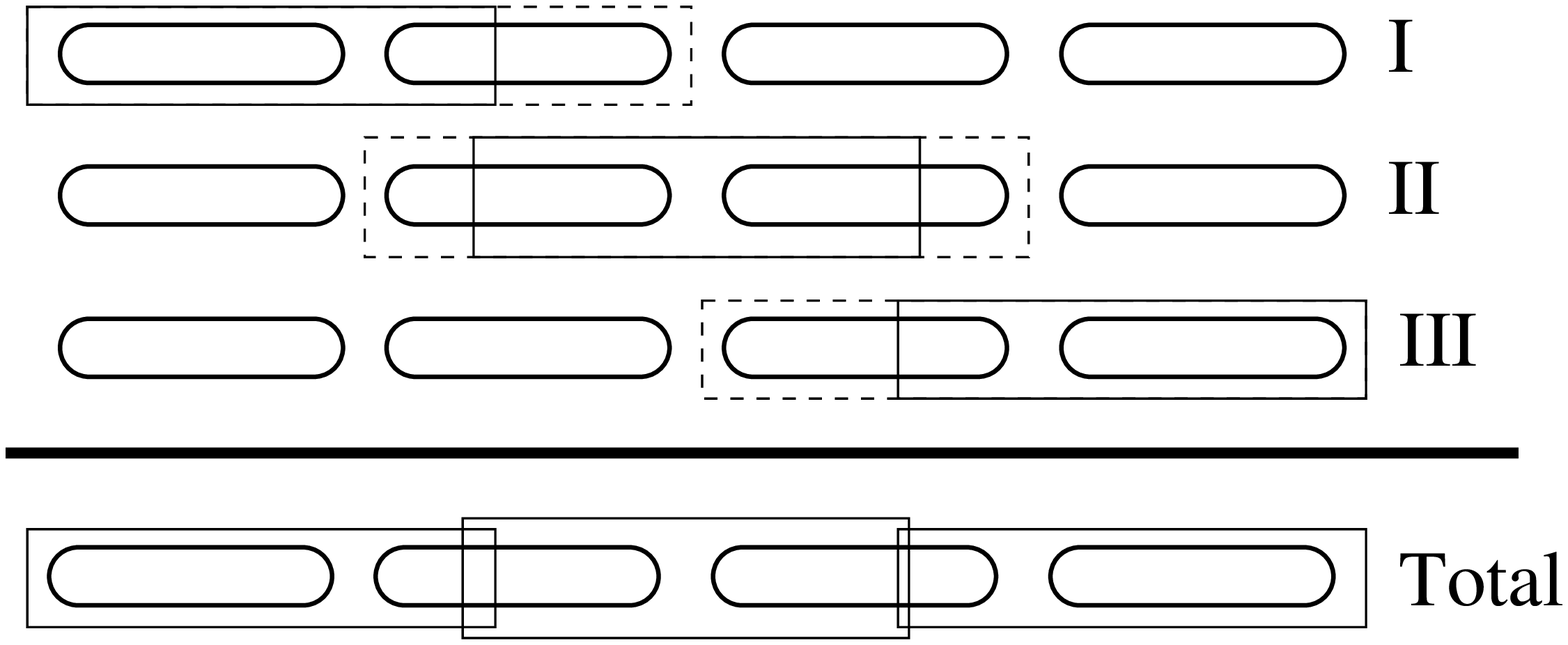}
\figlabel{\ref{fig:lrdm}}
\includegraphics{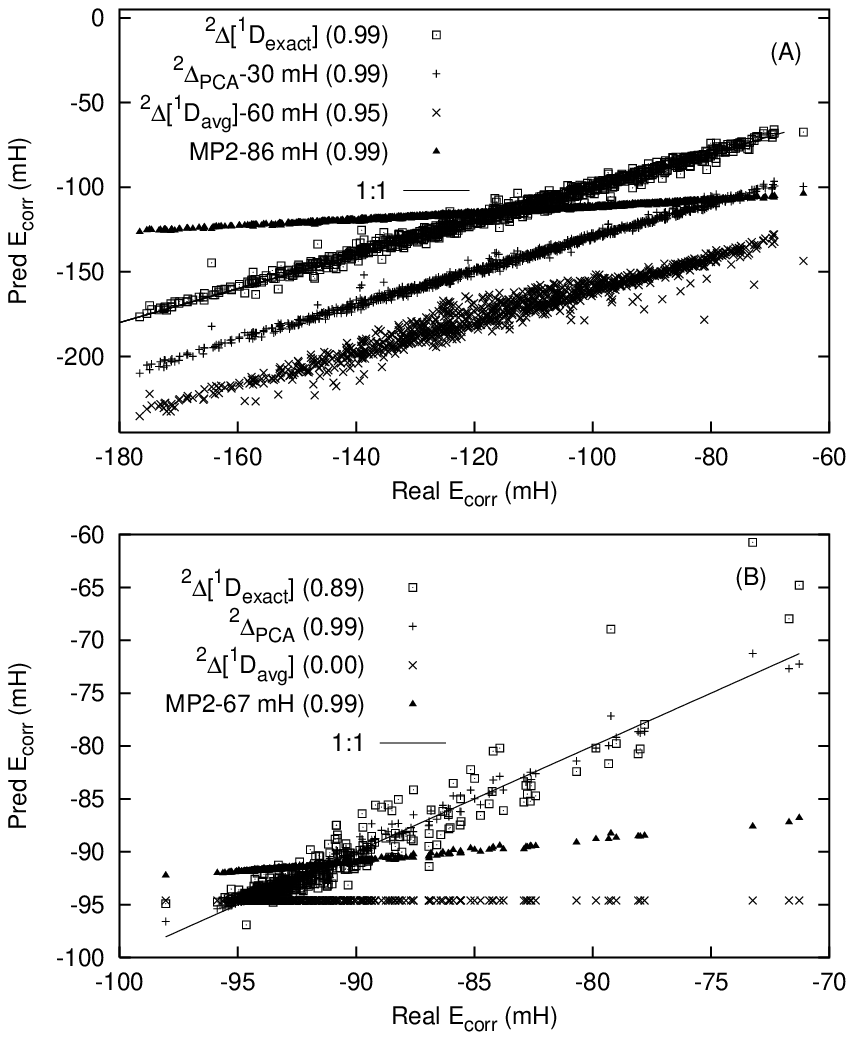}
\figlabel{\ref{fig:res1}}
\includegraphics{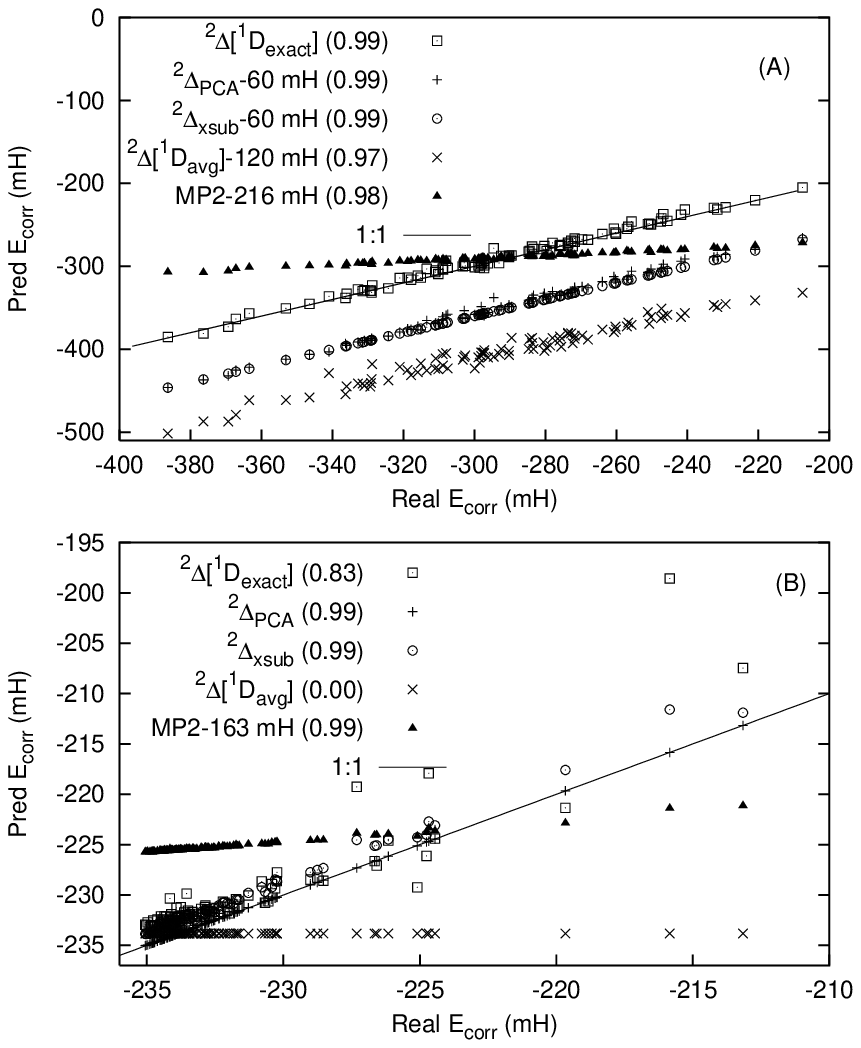}
\figlabel{\ref{fig:res2}}
\includegraphics{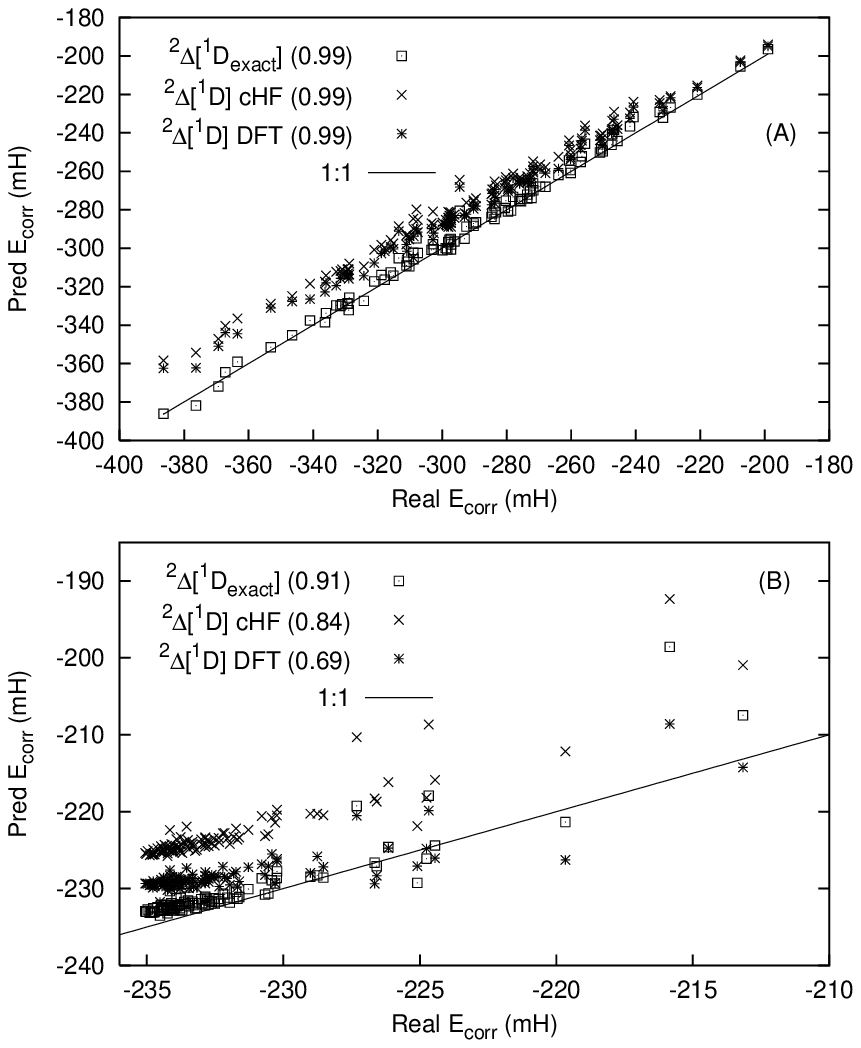}
\figlabel{\ref{fig:dft2}}
\includegraphics{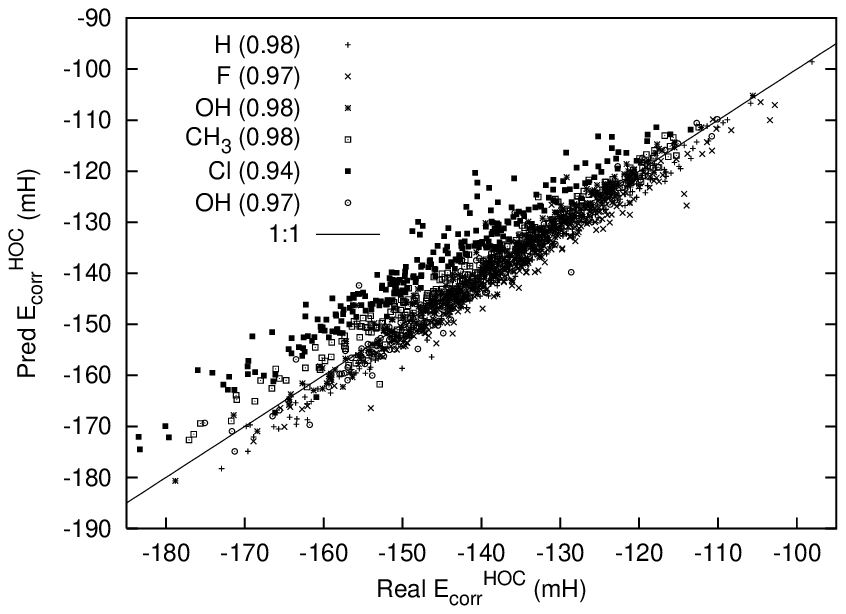}
\figlabel{\ref{fig:aldplot2}}

\end{document}